\documentclass[aps,prb,twocolumn,showpacs,superscriptaddress,groupedaddress]{revtex4-1}  
\pdfoutput=1
\usepackage{graphicx}  
\usepackage{dcolumn}   
\usepackage{bm}        
\usepackage{amssymb}   
\usepackage[usenames,dvipsnames]{xcolor}

\usepackage{amsmath,amsfonts,amssymb,amsthm}
\usepackage{latexsym}

\usepackage{tabularx}
\usepackage[figuresright]{rotating} 

\newcommand{\V}[1]{ \bm{#1}}
\newcommand{\SCH}{Schr\"{o}dinger }
\newcommand{\PoutT}{P_\text{out}' }
\newcommand{\diff}{\mathop{}\!d}
\newcommand{\PSIon}{$\text{Ps}^-$}
\newcommand{\bra}[1]{\left\langle #1 \right| }

\newcommand{\ket}[1]{\left| #1 \right\rangle }
\newcommand{\pick}{\lambda _\text{po}}

\newcommand{\braket}[2]{\left\langle #1 | #2 \right\rangle }
\newcommand{\Para}{\text{\emph{p}-Ps }}
\newcommand{\Orto}{\text{\emph{o}-Ps }}
\newcommand{\quotes}[1]{``#1''}
\newcommand\mydots{\ifmmode\ldots\else\makebox[1em][c]{.\hfil.\hfil.}\thinspace\fi}
\newcommand{\myref}[1]{(\ref{#1})}
\newcommand{\myfig}[1]{\ref{#1}}
\newcommand{\mysec}[1]{\ref{#1}}

\begin{document}

\title{Formal calculation of exchange effects on confined positronium}

\author{G. Marlotti Tanzi}
\email{giacomo.tanzi14@gmail.com}
\affiliation{Department of Physics ``Aldo Pontremoli'',
	Universit\`{a} degli Studi di Milano, via Celoria 16 I-20133 Milano, Italy}

\author{G.Consolati}
\affiliation{Department of Aerospace Science and Technology, Politecnico di Milano,
via La Masa 34 I-20156 Milano Italy}
\affiliation{INFN, sezione di Milano, via Celoria 16 I-20133 Milano, Italy}

\author{F. Castelli}
\affiliation{Department of Physics  ``Aldo Pontremoli'',
	Universit\`{a} degli Studi di Milano, via Celoria 16 I-20133 Milano, Italy}
\affiliation{INFN, sezione di Milano, via Celoria 16 I-20133 Milano, Italy}

\date{\today}

\begin{abstract}
Positronium atoms (Ps) are widely used as a probe to characterize
voids or vacancies in non-metallic materials, where Ps annihilation
lifetime is strongly modified by pickoff, depending on the size of
the trapping cavity and on the appropriate outer electron density.
The connection between these material characteristics and Ps annihilation
lifetimes is usually based on models that do not consider the requirements
of full electron indistinguishability, which must be taken into account
for a correct description of pickoff annihilation processes.
In this report we provide a formal theoretical framework in which exchange
effects between Ps and surrounding electrons are introduced in a natural
way, giving a clear and versatile picture of the various contributions to
the Ps pickoff annihilation. Moreover, our results provide a simple
explanation of the lowering of the contact density (the Ps-electron density
at the positron position) as a direct
consequence of the electrons indistinguishability, at variance with previous
interpretation based on spatial deformations of Ps wavefunction.
Calculations are performed within the "symmetry adapted perturbation
theory" approach, and the results are compared with experimental data
on Ps lifetimes of some polymers and molecular solids.
Finally, introducing suitable
approximations, we also recover early modeling and give a simple
interpretation of Ps properties in subnanometric voids.	
\end{abstract}

\maketitle


\section{Introduction}

In recent years the hydrogenlike bound state of an electron and a positron,
namely the positronium atom (Ps), has been extensively studied in the context
of structural analysis of porous materials.
In particular, positron annihilation lifetimes spectroscopy (PALS) is one of the few methods available to obtain information about sub-nanometric porous structures (i.e. defects, voids, cavities and free spaces in general) which may be present inside a sample \cite{jean2003principles}.

In condensed matter, Ps lifetimes result deeply different from the corresponding vacuum values, which depend only on its internal spin configuration. They are equal to $\tau_{2\gamma}=\lambda_{2\gamma}^{-1}=0.125\,\text{ns}$  and  $\tau_{3\gamma}=\lambda_{3\gamma}^{-1}=142\,\text{ns}$ for the singlet (\Para) and triplet (\Orto) state respectively, where $\lambda_{2\gamma}$ and $\lambda_{3\gamma}$ are the corresponding annihilation rates.

As a matter of fact, a complete theory of Ps formation and annihilation inside matter is needed in order to
extract useful information about the medium itself from PALS data.
At present, this is usually achieved through an approximate one or two-body description of the so called \textit{pickoff} process, i.e. the possibility for the positron to annihilate with an electron of the surroundings, different from that to which is bound in a Ps atom.

The most used one-body models describing Ps inside materials with small cavities are based on the Tao-Eldrup (TE) approach \cite{Tao,Eldrup81}, which relates
pick-off annihilation rates $\pick$ to pore sizes by considering Ps as a single quantum particle trapped inside an infinite
potential well.
At the state of the art, these models have been greatly extended to describe various cavity geometries
and temperature effects \cite{Dutta,RTE}.

Historically introduced as a natural extension of TE, two-body models describe also the internal structure of Ps by considering separate degrees of freedom for the positron and the electron \cite{Stepanov,Tanzi,Bug2}.
Also, fully \textit{ab initio} treatments of a two particle bound system inside a host material can in principle be done \cite{Puska2}, but they are usually avoided given the huge computational efforts required.

In this context, it has long been assumed in literature that Ps interaction with external electrons can be described as a small perturbation.
This assumption is implicitly at the basis of every one-body and two-body models, where the outer electronic environment accounts for pickoff annihilation without modifying the nature and the form of Ps as a bound state.
In particular, it is believed that a two-body approach is the simplest one capable of describing any variation of the \textit{intrinsic} relative contact density parameter $k_r$, defined as the probability of finding the Ps-electron at the positron position in units of the vacuum value $k_0 = 1/8\pi a_0^3$ ($a_0$ being the Bohr radius).
We used the term \quotes{intrinsic} to differentiate this quantity from the analogous \quotes{total} contact density parameter usually found in positron physics, which is proportional to the probability of finding \textit{any} electron at the positron position.

However, the validity of a theoretical treatment in which the Ps is seen as a separate \quotes{entity} and where the Ps-electron is somehow privileged with respect to outer electrons must be questioned against the requirement of full electron indistinguishability, especially given its direct relation to the pickoff annihilation.
Hints on the possibility of treating the Ps-electron in a different way come from PALS experiments in materials showing different lifetime signatures, being this a direct evidence of the presence of a statistical mixture of different Ps states. In literature, these are usually interpreted in terms of \Para and \emph{o}-Ps. Such a distinction requires the identification of a specific electron, whose spin couples with the positron in a specific singlet or triplet configuration. On the contrary, complete electron indistinguishability is evident in materials and compounds exhibiting a single lifetime component (the simplest example being \PSIon).
Furthermore, we note that a Ps-like component in the spatial (or momentum) part of the wavefunction describing a positron in matter does not necessarily imply the presence of different Ps states (for example, a delocalized Ps may be present as a single superposition of singlet/triplet states).

In this work we analyze in detail this problem, providing a theoretical framework in which electron indistinguishability can be introduced in a natural way, yet preserving the concept of para/ortho Ps. We will focus on a particular aspect of this problem, that we call \quotes{over-counting}, which plays an important role in the study of the annihilation process of Ps in cavities.

Finally, our picture will also provide a simple explanation for the well known phenomenon of the lowering of the intrinsic contact density with respect to its vacuum value, as it is found in many solid materials. By connecting this phenomenon to electron indistinguishability, we will show how it is by no means related to a spatial deformation of Ps wavefunction, as previously believed.

\section{The over-counting problem} \label{overcounting}
The most common set of equations used to describe \Orto and \Para annihilation rates in porous matter, respectively $\lambda_t$ and $\lambda_s$, is given in literature by \cite{Dupasq2}:
\begin{subequations}\label{DupasqRates}
	\begin{align}
	\lambda_t &= k_r \lambda_{3\gamma} + \pick \label{ortoDupas} \\
	\lambda_s &= k_r \lambda_{2\gamma} + \pick \label{paraDupas}
	\end{align}
\end{subequations}
For long time it has been thought of $k_r$, the usual relative contact density, as an intrinsic property of the confined Ps, whereas the term $\pick$, which is identical in both Eq.~\myref{ortoDupas} and \myref{paraDupas}, was associated to the pickoff annihilation process with outer electrons.
Given that pickoff is by nature a \textit{surface} process, in every model $\pick$ was assumed to depend on a geometrical probability, commonly denoted by $P_\text{out}$, of finding Ps outside the free-space (\textit{inner}) region defining the cavity:
\begin{equation}
\pick = P_\text{out} \lambda_b
\end{equation}
where $\lambda_b$ is a suitable bulk annihilation rate.
It has become a common practice to fix $ \lambda_b$ to the weighted
average of singlet and triplet decay rates $\bar{\lambda}= \frac{1}{4}\lambda_{2\gamma}+ \frac{3}{4}\lambda_{3\gamma}=2.01 [\text{ns}]^{-1} $, following a prescription originally due to TE:
\begin{equation}\label{pickofffirst}
\pick = P_\text{out} \bar{\lambda}
\end{equation}
Being independent of the electronic properties of the surrounding medium, such an assumption must be regarded as an effective approximation, which holds provided that the geometrical parameters of the model are consequently chosen to fit the correct pickoff annihilation in real systems.

It came to our attention that there are many different hypothesis about the proper way of treating Ps in the \textit{inner} and \textit{surface} regions.
In many works (for example \cite{goworek1998positronium,ito1999extension,jean2013perspective}) Ps, described as a single particle with $k_r =1$, is considered affected on the same foot by both intrinsic and pickoff annihilations in the outer part of the cavity:
\begin{equation}\label{overcountedRATES}
\begin{aligned}
\lambda_t &= \lambda_{3\gamma} + P_\text{out} \lambda_b\\
\lambda_s &= \lambda_{2\gamma} + P_\text{out} \lambda_b\\
\end{aligned}
\end{equation}

On the other hand, a few one-particle models (to our knowledge this was done only in \cite{RTE,goworek2002comments,zaleski2003electron}) completely differentiate the \textit{inner} and \textit{surface} description of Ps.
In these, Ps annihilates with its intrinsic vacuum annihilation rate only in the \textit{inner} part of the cavity, whereas the \textit{surface} region is dominated by pickoff.
Following Goworek \cite{goworek2002comments}, Eqs.~\myref{overcountedRATES} are written in this picture as:
\begin{equation}\label{NOovercountedRATES}
\begin{aligned}
\lambda_t &= (1-P_\text{out}) \lambda_{3\gamma} + P_\text{out} \lambda_b\\
\lambda_s &= (1-P_\text{out}) \lambda_{2\gamma} + P_\text{out} \lambda_b\\
\end{aligned}
\end{equation}
being $(1-P_\text{out}) = P_\text{in}$ the probability of finding Ps in the \textit{inner} free-space region.
Remarkably, a direct comparison between Eqs.~\myref{DupasqRates} and Eqs.~\myref{NOovercountedRATES} show that the latter have by construction an intrinsic relative contact density $k_r =  P_\text{in}$ lower than unity.
Surprisingly enough, to our knowledge, this important connection has gone unnoticed by the authors and by the positronium community until now.
In~\cite{goworek2002comments} this was due to an erroneous interpretation of the contact density, while in~\cite{RTE} no considerations about the contact density were done at all.

Finally, a somehow intermediate situation is found in all two-particle models (for example \cite{stepanov2012positronium,Tanzi,Tanzi2}), where pickoff annihilation is proportional to the probability $P_\text{out}^+$ of having the \textit{positron} outside the cavity
\begin{equation}
\begin{aligned}
\pick  &=\lambda_b P_\text{out}^+\\
\end{aligned}
\end{equation}
which is somehow similar to $P_\text{out}$.
In these models, intrinsic annihilation is assumed to take place only in the region allowed to the Ps-electron that, analogously to Eqs~\myref{overcountedRATES} and \myref{NOovercountedRATES}, can be either extended to the whole space \cite{stepanov2012positronium} or limited to the \textit{inner} cavity (if Ps-electron is striclty confined, like in \cite{Tanzi}).

In our view, all these different approaches are due to a general lack of clarity about the meaning of terms appearing in Eqs.~\myref{DupasqRates}.
In particular, the fact that both the expressions for $\lambda_t$ and $\lambda_s$ in Eqs.~\myref{DupasqRates} have the \textit{same structure}, has been erroneously interpreted by some as the prove that \Orto and \Para are affected by the \textit{same pickoff annihilation rate}.
In other words, it is assumed that a particular spin configuration of the Ps-electron does not affect in any way the pickoff annihilation behavior of Ps-positron in the outer layer.
As a direct consequence, the pickoff process was exclusively linked to the term $\pick$ in Eqs.~\myref{DupasqRates}, while $k_r$ was associated to possible modifications of the internal spatial structure of Ps wavefunction.

\begin{figure}
	\centering
	\includegraphics[width=\linewidth]{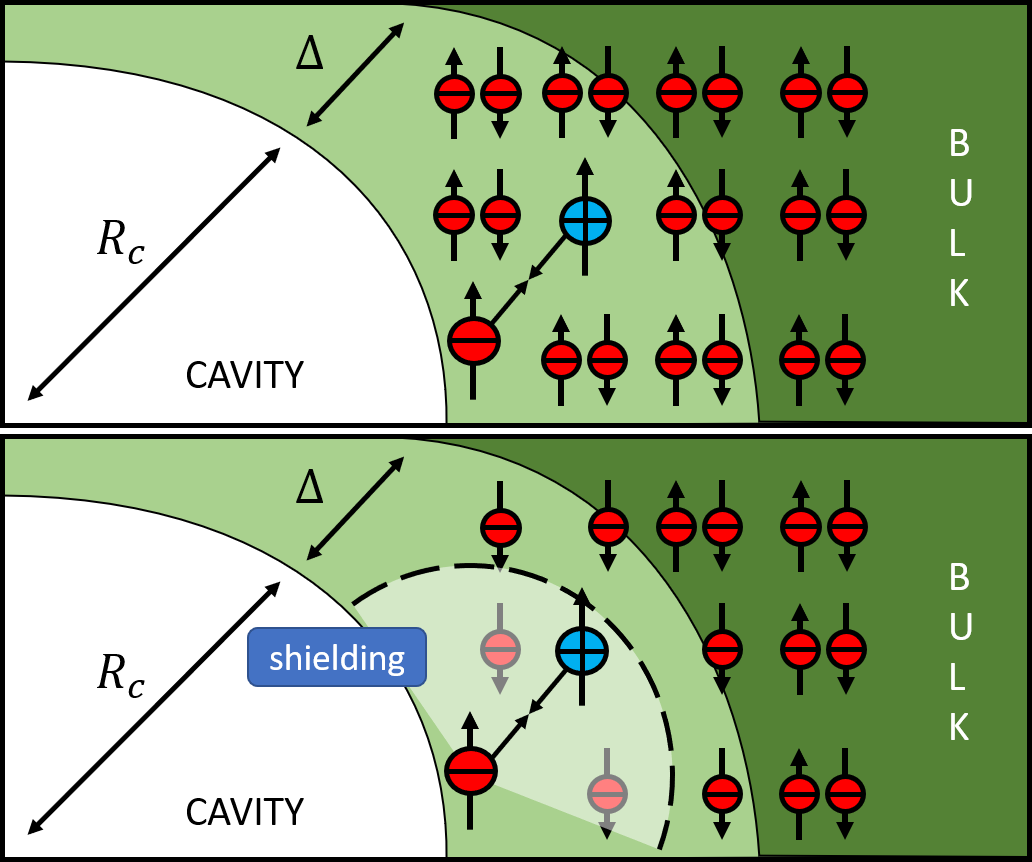}
	\caption{Effect of electron shielding on positron annihilation. $R_c$ and $\Delta$ are commonly used symbols delimiting the \textit{inner} and \textit{surface} region respectively.
		Top: without shielding, the positron is free to annihilate with outer electrons of any spin configuration. Bottom: if shielding is considered, the positron will most likely annihilate with electrons having opposite spin with respect to Ps-electron. }
	\label{fig:shielding}
\end{figure}

In this picture, no \quotes{shielding} effect due to exchange correlation effects (Pauli exclusion principle) is ascribed to the Ps-electron. Hence the positron is free to annihilate with all surrounding electrons, independently from their spin, with a consequent \textit{over-counting} of annihilation processes inside the \textit{surface} region (as sketched in Fig.~\myfig{fig:shielding}).
Surprisingly, this no-shielding assumption was neither fully justified nor properly discussed from a theoretical point of view.
The possibility of having different pickoff annihilation rates for \Orto and \Para due to spin exchange was only noted, to our knowledge, by Mogensen and Eldrup in 1977 \cite{mogensen1977positronium}, but never further investigated.
Anyway, the lack of such a discussion represents a minor problem to the positronium community since the over-counting has a negligible effect on the total annihilation rate of the \Orto system (i.e. the easily measurable long life component of PALS spectra), where $\pick \gg \lambda_{3\gamma}$. The same is not true for \Para, where pickoff and intrinsic annihilation rates may be comparable.

The question of whether this over-counting is legitimate or not must be answered in the framework of many-body quantum mechanics.
This will be discussed in detail in the following sections, where we will show how the pickoff annihilation rate \textit{is indeed different} for \Orto and \Para.
Here, we just note that this statement is not in contrast with Eqs.~\myref{DupasqRates} as long as one realizes that they can be written as:
\begin{subequations}\label{correct_interpretation}
	\begin{align}
	\lambda_t &= \lambda_{3\gamma} + \left[  (k_r-1) \lambda_{3\gamma} + \pick \right]  \\
	\lambda_s &= \lambda_{2\gamma} + \left[  (k_r-1) \lambda_{2\gamma} + \pick \right]
	\end{align}
\end{subequations}
where the term in square brackets can be interpreted as the overall contribution to the annihilation due to the external electrons, i.e. the pickoff.
This kind of formula has exactly the same form of the one that will be derived from the theory developed in the following sections.

\section{Exchange perturbation theories}

 The detailed quantum state of an electron-positron pair inside a cavity is extremely complex.
 Whereas in the inner part of the free-space region it will resemble an isolate Ps bound state, in the outer part it will fade into a \quotes{spur state} (sometimes called quasi-Ps) \cite{mogensen1974spur} of a positron interacting with the full many-body environment.
 The main difficulty arises from the fact that in the first scenario one has a separate Ps-electron, while in the other complete electron indistinguishability must be taken into account.

The formulation of a theoretical treatment apt to describe the transition between these two limiting situations is an old problem in both physics and chemistry.
There are many systems (e.g. atoms in molecules) wherein individual components are clearly identifiable and, in the non-interacting picture, may be described by an asymptotic-free hamiltonian $H_0 = H_A + H_B$ where electrons are arbitrarily assigned to different subsystems $A$ and $B$. In this asymptotic picture, the ground state wavefunction $\psi^{(0)} =\psi_A\psi_B $ can be written in a factored form and does not need to be fully antisymmetric.
 Since extramolecular interactions $V_{AB}=H- H_0$ in these systems are often small compared with the low-lying intramolecular (or intraatomic or intraionic) level spacings, some sort of perturbative treatment based upon noninteracting components is suggested \cite{kleinExchangePerturbationTheories}.

To extend the treatment overcoming the antisymmetry problem, since the 1960s a vast class of symmetry-adapted perturbation theory (SAPT) were proposed \cite{szalewicz2005intermolecular}, and this is the reference theoretical framework in which our theory is going to be developed.
An accurate review of SAPT is beyond the scope of the present discussion and can be found in \cite{klein2010composite}.
In particular, in all SAPT formulations, the first order correction to the energy of the composite systems reads:
\begin{equation}\label{SAPT1}
\begin{aligned}
E^{(1)} &= \frac{\bra{\psi^{(0)}}V_{AB}\ket{\mathcal{A}\psi^{(0)}}}{\braket{\psi^{(0)}}{\mathcal{A}\psi^{(0)}}} \\
\end{aligned}
\end{equation}
Here, $\mathcal{A}$ is an intermolecular antisymmetrizer operator, defined as\footnote[2]{\protect With this definition, $\mathcal{A}$ is idempotent, i.e. $\mathcal{A}^2 = \mathcal{A}$.}:
\begin{equation}	\label{antisymmetrizer}
\mathcal{A}=\frac{1}{N!}\sum_p(-1)^p P
\end{equation}
where $P$ represents a permutation operator of $N$ electrons, while $(-1)^p$ stands for the parity of the permutation.
The factor $\braket{\psi^{(0)}}{\mathcal{A}\psi^{(0)}}= \braket{\psi^{(0)}\mathcal{A}}{\mathcal{A}\psi^{(0)}}$ at the denominator of Eq.~\myref{SAPT1} explicitly takes into account the so-called intermediate-normalization condition \cite{klein2010composite}.

\section{The Ps-environment system}\label{PsPerturbationInducedByExternalElectrons}
With the aim of applying SAPT methods to our problem, we proceed towards a suitable setting up of the Ps-environment system.
The most general Hamiltonian of a system composed of a Ps atom interacting with an $N$-electron environment can be written as a sum of a free Ps Hamiltonian $\hat{H}^{(0)}_\text{Ps}$, the Hamiltonian of the material $\hat{H}_b$ and an interaction potential acting between these two subsystems.
Considering only Coulomb interactions and neglecting atomic nuclei, which are not involved in the annihilation process, we write $\hat{H}$ as:
\begin{widetext}
\begin{equation}\label{Hamiltonian0}
\begin{aligned}
\hat{H} &= 	\hat{H}^{(0)}_\text{Ps}(\V r _p,\V r _e) + 	\hat{H}_b(\V r _1,\V r _2,\cdots,\V r _N) + \sum_{i=1}^N \left[ \hat{V}_\text{C}(\V r _e,\V r _i)-\hat{V}_\text{C}(\V r _p,\V r _i)\right]\\
&\equiv\hat{H}^{(0)}_\text{Ps}(\V r _p,\V r _e) + 	\hat{H}_b(\V r _1,\V r _2,\cdots,\V r _N) + \sum_{i=1}^N \hat{V}_\text{int}(\V r _p,\V r _e,\V r _i)\\
\end{aligned}
\end{equation}
\end{widetext}
where $\hat{V}_\text{C}(\V r _x,\V r _y)$ is the Coulomb potential between the particles $x$ and $y$. In the following we will denote $p=(\V r _p , \sigma_p)$ and $e=(\V r _e , \sigma_e)$ the spin-spatial coordinates of the Ps positron and electron, respectively, while numbers refer to other electrons for convenience.

From the success of many theoretical models describing Ps in porous materials, we know that the overall effect of interactions can be well described by an effective potential $\hat{V}_\text{eff}(\V r _p,\V r _e)$ which acts only on the spatial coordinates of the Ps atom as a whole.
Despite this potential can be found in different formulations in literature, the most important feature they all share is the confining effect. As an example, in the TE model this potential is taken as an infinite quantum well $\hat{V}_\text{eff}(\V r _p,\V r _e) = \hat{V}_{\infty}(\V  R)$ acting on Ps center of mass $\V  R$.
Hence, it is convenient to include this potential in the definition of the Ps hamiltonian, so that Eq.~\myref{Hamiltonian0} can be reformulated as:
\begin{widetext}
\begin{equation}\label{Hamiltonian}
\begin{aligned}
\hat{H} &= 	\hat{H}_\text{Ps}(\V r _p,\V r _e) + 	\hat{H}_b(\V r _1,\V r _2,\cdots,\V r _N) + \left[\sum_{i=1}^N \hat{V}_\text{int}(\V r _p,\V r _e,\V r _i) - \hat{V}_\text{eff}(\V r _p,\V r _e)\right]\\
&\equiv \hat{H}_\text{Ps}(\V r _p,\V r _e) + 	\hat{H}_b(\V r _1,\V r _2,\cdots,\V r _N) + \hat{V}\\
\end{aligned}
\end{equation}
\end{widetext}
where $\hat{H}_\text{Ps} = \hat{H}^{(0)}_\text{Ps} + \hat{V}_\text{eff} $ is the effective Ps hamiltonian.

In the framework of a perturbative approach, by neglecting the interaction potential $\hat{V}$, the Hamiltonian in Eq.~\myref{Hamiltonian} becomes separable and its ground state will be the product of a Ps wavefunction $\Psi_{jm}$ times the (antisymmetric) ground state $\phi$ of the $N$-electron system:
\begin{equation}\label{zeroGroundState}
\begin{aligned}
\psi^{(0)}_{jm}(p,e,1,\cdots,N) &= \Psi_{jm}(p,e)\phi(1,2,\cdots,N)\\
\end{aligned}
\end{equation}
where $j,m$ are the Ps spin $|S|^2$ and spin projection $S_z$ quantum numbers ($j=1$ for \Orto and $j=0$ for \Para).
The wavefunction of the system $\psi^{(0)}(p,e;1,\mydots,N)$ is by construction antisymmetric with respect to the exchange of two electrons in $1,\mydots,N$ since
\begin{equation}
\phi(1\cdots i \cdots j \cdots N) = - \phi(1\cdots j\cdots i \cdots N) \qquad\forall i,j
\end{equation}
for every $j,i$, but it is not antisymmetric with respect to the exchange with Ps electron.

To ease the notation in the following discussion, we introduce now some quantities which are usually well-known.
The external electron density is connected to the square modulus of the $N$-electron normalized wavefunction and it is defined as
\begin{equation}\label{electronDensity}
\begin{aligned}
n(\V r) = &  N \sum_{\sigma_1}\int |\phi(\V r,\sigma_1,2,\cdots,N)|^2\diff 2 \mydots \diff N\\
\end{aligned}
\end{equation}
Here and in the following we use the compact notation $\int \diff i = \sum_{\sigma_i}\int \diff^3 r_i $ to represent both spin summation and spatial integration. Moreover, for simplicity in writing we may omit to specify integration variables $\diff i$ and domain when these are evident, as often done in such a kind of calculations.

A commonly used concept in many body physics is that of reduced density matrices (RDM), which offers a convenient way of describing the internal structure of a many body system of $N$ indistinguishable particles without the complete knowledge of its wavefunction.
The term \quotes{reduced} refers to the fact that attention is focused on a reduced number of coordinates, being the density matrix of the total system averaged over
all the others.
The simplest RDM is the one body reduced density matrix (1RDM), which is defined as:
\begin{equation}\label{oneBodyRedM}
\begin{aligned}
\Gamma^{(1)}(x;y) = & N\int \phi(x,2,\cdots,N)\phi^*(y,2,\cdots,N) \diff 2 \cdots \diff N\\
\end{aligned}
\end{equation}
where, as stated before, $x$ and $y$ denotes the couple $(\V r _x , \sigma_x)$ and $(\V r _y , \sigma_y)$.
The 1RDM has in principle 4 components $\Gamma^{(1)}_{\uparrow\uparrow}$, $\Gamma^{(1)}_{\uparrow\downarrow}$, $\Gamma^{(1)}_{\downarrow\uparrow}$ and $\Gamma^{(1)}_{\downarrow\downarrow}$ resulting from expansion in a complete set of spin functions:
\begin{equation}\label{oneBodyRedMproperties1}
\begin{aligned}
\Gamma^{(1)}(x;y) = & \sum_{ij}\Gamma^{(1)}_{ij}(\V r _x;\V r _y)s_i(\sigma_x)s_j^*(\sigma_y)\\
\end{aligned}
\end{equation}
where $i$ and $j$ may represent $\uparrow$ or $\downarrow$ spin states.
Furthermore, we can define the spatial 1RDM by integrating $\Gamma^{(1)}$ over the spin variables:
\begin{equation}\label{oneBodyRedMproperties1b}
\begin{aligned}
\Gamma^{(1)}(\V r _x;\V r _y) = & \sum_{\sigma_x,\sigma_y}\sum_{ij}\Gamma^{(1)}_{ij}(\V r _x;\V r _y)s_i(\sigma_x)s_j(\sigma_y)\\
\end{aligned}
\end{equation}
In general, if no spin mixing potential appears in the hamiltonian of the bulk system as assumed here, the wavefunction $\phi$ is an eigenstate of $S_z$ and the two spin channels decouple, so that $\Gamma^{(1)}_{\uparrow\downarrow} = \Gamma^{(1)}_{\downarrow\uparrow} = 0$ and \cite{davidson2012reduced}:
\begin{equation}\label{oneBodyRedMproperties2}
\begin{aligned}
\Gamma^{(1)}(\V r _x;\V r _y)  = & \Gamma^{(1)}_{\uparrow\uparrow}(\V r _x;\V r _y) +\Gamma^{(1)}_{\downarrow\downarrow}(\V r _x;\V r _y)
\end{aligned}
\end{equation}
Finally, the diagonal part of the spatial 1RDM is just the electron density defined in Eq.~\myref{electronDensity}:
\begin{equation}\label{oneBodyRedMproperties}
\begin{aligned}
&n(\V r ) = \Gamma^{(1)}(\V r ;\V r ) = n_\uparrow(\V r) + n_\downarrow(\V r)\\
\end{aligned}
\end{equation}
where $ n_\uparrow$($n_\downarrow$) is the local spin up(down) density.

Another useful quantity is the two body reduced density matrix (2RDM), defined as:
\begin{equation}\label{twoBodyRedM}
\begin{aligned}
&\Gamma^{(2)}( x,x';y,y') \\
&\quad= \binom{N}{2}\int \phi(x,x',3,\cdots,N)\phi^*(y,y',3,\cdots,N) \diff 3 \cdots \diff N \\
\end{aligned}
\end{equation}
which also can be expanded over a complete set of spin functions, with a total of 16 components:
\begin{equation}\label{twoBodyRedMonSpin}
\begin{aligned}
\Gamma^{(2)}( x,x';y,y') &=\sum_{ij,i'j'}\Gamma^{(2)}_{iji'j'}( \V r _x,\V r _ {x'};\V r _ y,\V r _ {y'})\\
& \qquad \times s_{i}(\sigma_{x})s_{j}(\sigma_{x'})s_{i'}^*(\sigma_y)s_{j'}^*(\sigma_{y'}) \\
\end{aligned}
\end{equation}
As for 1RDM, a spatial 2RDM is introduced by integrating $\Gamma^{(2)}$ over the spin variables $\sigma_{x}$, $\sigma_{x'}$, $\sigma_y$ and $\sigma_{y'}$.
The diagonal part of the 2RDM, $\Gamma^{(2)}( \V r _ x,\V r _y;\V r _ x,\V r _ y) = P(\V r _ x,\V r _ y)$, is the pair distribution function, proportional to the conditional probability of having an electron in $\V r _ y$ given another one in $\V r _ x$.
Since the correlation between two electrons vanishes at long distances, in this limit is well known that $P(\V r _ x,\V r _ y)$ satisfies the condition:
\begin{equation}\label{2RDMlimit_condition}
P(\V r _ x,\V r _ y) \approx n(\V r _ x)n(\V r _ y) \qquad\text{when}\qquad|\V r _ x -\V r _ y| \to \infty
\end{equation}
On the other hand, the probability of having two electrons very close to each other is strongly suppressed in real systems by both the Pauli exclusion principle (if they have the same spin) and by the strong Coulomb repulsion.

When Ps approaches the external electronic system, its wavefunction will begin to \quotes{overlap} with the system's one and exchange correlation effects must be considered.
In this sense it is useful to quantify this overlap by introducing a suitable parameter $S$ with the following definition, whose special formulation will become clear in the next section:
\begin{equation}\label{S}
\begin{aligned}
S &= \sum_{i=1}^{N} \int \Psi^*_{jm}(p,e)\phi^*(i,2,\cdots,N)\Psi_{jm}(p,i)\phi(e,2,\cdots,N)\\
&= N \int \Psi^*_{jm}(p,e)\phi^*(1,2,\cdots,N)\Psi_{jm}(p,1)\phi(e,2,\cdots,N)\\
&=\int  \Psi^*_{jm}(p,e)\Psi_{jm}(p,1)\Gamma^{(1)}(e;1) \\
\end{aligned}
\end{equation}
where we used the antisymmetry properties of $\phi$.
Assuming a Ps atom confined \textit{a priori} in a certain free-space region (cavity) means that the interaction with the external electrons will take place only in a limited \textit{surface} domain, so that the support of integral in Eq.~\myref{S}, hence the overlap, is small by construction.

\section{Perturbative approach to annihilation rate}

As pointed out by many authors~\cite{igarashi2003inseparable}, the QED phenomenon of annihilation can be described in a simpler way through the introduction of an effective absorption potential $-i\hbar \hat{\lambda}/2$ in the ordinary time-dependent \SCH equation of the quantum mechanical system under examination, where $\hat{\lambda}$ is a suitable loss rate operator \footnote[3]{The factor $1/2$ arises from the fact that the general definition is given in terms of the probability density $ \protect\frac{\diff}{\diff t}|\psi|^2 = - \lambda |\psi|^2$, where as usual $\lambda = \tau^{-1}$ is the inverse of the lifetime.}.
Being imaginary, this potential leads to an exponential decay of the positron (positronium) wavefunction, which accounts for particle loss and whose rate can be determined via PALS experiments.
This peculiar representation of the annihilation process makes possible its description in terms of imaginary part of the energy of the system.
In particular, the first order correction to the annihilation rate can thus be derived from the (imaginary part of) first order correction to energy.
This correction can in turn be calculated in SAPT framework using Eq.~\myref{SAPT1}, with just the knowledge of the unperturbed ground state of the system
\footnote[4]{This is not restricted to SAPT but it is also true for standard RS perturbation theory. It is a well-known fact that the first order correction in energy is given by $E^{(1)} = \protect \bra{\psi^{(0)}}V\protect \ket{\psi^{(0)}}$ and does not require the calculation of the first order correction to the ground state $\protect \ket{\psi^{(1)}}$.}.

To take advantage of SAPT description, we need to split the loss rate operator into a \quotes{intramolecular} part $\hat{\lambda}_e$, related only to the \textit{intrinsic} annihilation of the positron with the Ps-electron $e$, and an \quotes{extramolecular} part $\sum_{i=1}^N \hat{\lambda}_i$, related to pickoff annihilations coming from the other $N$ electrons. The Hamiltonian operator~\myref{Hamiltonian} becomes now:
\begin{equation}\label{Hamiltonian1}
\begin{aligned}
\hat{H} =& 	\hat{H}_\text{Ps}(\V r _p,\V r _e) -i\frac{\hbar}{2} \hat{\lambda}_e +	\hat{H}_b(\V r _1,\V r _2,\cdots,\V r _N) + \\
&  \qquad\qquad\qquad\qquad\qquad+\left[\hat{V} -i\frac{\hbar}{2}\sum_{i=1}^N   \hat{\lambda}_i\right]\\
\end{aligned}
\end{equation}
The explicit expression of the annihilation operator in this picture is given by\cite{Dupasq2}:
\begin{equation}\label{annihilationoperator}
\hat{\lambda}_i = 8\pi a_0^3 \delta^3(\V r _p-\V r _i) \left[\frac{1-\Sigma_{p,i}}{2}\lambda_{2\gamma}+\frac{1+\Sigma_{p,i}}{2}\lambda_{3\gamma}\right]
\end{equation}
where $8\pi a_0^3$ is the inverse contact density of unperturbed positronium, $\V r _p$ and $\V r _i$ are positron and electrons coordinates, respectively, and $\Sigma_{p,i}$ is the spin exchange operator.
In this approximation  $\hat{\lambda}$ is basically a \quotes{contact operator}, being a linear combination of delta functions of the electron-positron distance. The spin exchange operator $\Sigma$ guarantees that the antisymmetric spin state annihilates via $2\gamma$ emission while the symmetric spin state via $3\gamma$ emission.
It's easy to see that this form of $\hat{\lambda}_i$ gives the correct annihilation rates for \Para and \Orto states in vacuum.
It is now straightforward to calculate the total annihilation rate for Ps
\begin{equation}
\lambda = \lambda^{(0)} +\lambda^{(1)}
\end{equation}
where the zero-order term is simply the \textit{intrinsic} annihilation rate, which does not depends on external electrons
\begin{equation}
\begin{aligned}
	\lambda^{(0)} &= \bra{\psi^{(0)}_{jm}}\hat{\lambda}_e\ket{\psi^{(0)}_{jm}}\\
	 &\equiv \bra{\Psi_{jm}}\hat{\lambda}_e\ket{\Psi_{jm}}\\
	 &=8\pi a_0^3\int \left|\Psi_{jm}(p,p)\right|^2 \times\left\lbrace\begin{aligned}
	 &\lambda_{2\gamma} &  \text{if $j=0$ (\Para)}\\
	 &\lambda_{3\gamma} &  \text{if $j=1$ (\Orto)}\\
	 \end{aligned}\right. \\
\end{aligned}
\end{equation}

The first-order correction represents the pickoff contribution and is determined from Eq.~\myref{SAPT1}:
\begin{equation}\label{lambdaSAPTcorrection}
\begin{aligned}
\lambda^{(1)} &= \frac{\bra{\psi^{(0)}_{jm}}\sum_{i=1}^N   \hat{\lambda}_i\ket{\mathcal{A}\psi^{(0)}_{jm}}}{\braket{\psi^{(0)}_{jm}}{\mathcal{A}\psi^{(0)}_{jm}}} \\
\end{aligned}
\end{equation}
Explicitly, using the definition given in Eq.~\protect\myref{antisymmetrizer}, we have:
\begin{equation}\label{APsi0normalized}
\begin{aligned}
\ket{\mathcal{A}\psi^{(0)}_{jm}}=&\frac{1}{(N+1)}\left[\Psi_{jm}(p,e)\phi(1,2,\cdots,N)\right.\\
-&\sum_{i=1}^N\left.\Psi_{jm}(p,i)\phi(1,\cdots,i-1,e,i+1,\cdots,N)\right]\\
\braket{\psi^{(0)}_{jm}}{\mathcal{A}\psi^{(0)}_{jm}}&= \frac{\mathcal{N}}{N+1}\\
\end{aligned}
\end{equation}
where the factor $(N+1)$ at the denominator is the total number of extra permutations of the Ps electron, while $\mathcal{N}$ depends only on the overlap $S$:
\begin{equation}\label{T0}
\begin{aligned}
\mathcal{N}&=1-\int  \Psi_{jm}^*(p,e)\Psi_{jm}(p,1)\Gamma^{(1)}(e;1) \\
&= 1-S\\
\end{aligned}
\end{equation}
From now on, for the sake of definiteness, we will focus on a particular component of \Orto. Hence we fix $\lbrace jm\rbrace=\lbrace 11\rbrace$ for simplicity, but analogous calculation can be done for any Ps state.
Taking $\mathcal{N}$ on the left side of Eq.~\myref{lambdaSAPTcorrection}, this last one can be written as:
\begin{equation}\label{generalexpression}
\begin{aligned}
&\mathcal{N}\lambda^{(1)}= \\
&\quad= N\int  \Psi^*(p,e)\phi^*(1,2,\cdots,N) \hat{\lambda}_1\Psi(p,e)\phi(1,2,\cdots,N)\\
&\quad-N\int \Psi^*(p,e)\phi^*(1,2,\cdots,N)\hat{\lambda}_1\Psi(p,1)\phi(e,2,\cdots,N)\\
&\quad-\binom{N}{2}2\int \Psi^*(p,e)\phi^*(1,2,\cdots,N)\hat{\lambda}_1\times\\
&\quad\qquad\qquad\qquad\qquad\qquad\qquad \times\Psi(p,2)\phi(1,e,\cdots,N)\\
\end{aligned}
\end{equation}
where we have used the antisymmetric property of $\phi$ to group together terms corresponding to the same contribution.
Eq.~\myref{generalexpression} shows that the overall correction to the annihilation rate is the sum of 3 different terms:
\begin{equation}\label{totalannihilation}
\mathcal{N}\lambda^{(1)} = \pick+\lambda_\text{ex}+\lambda_\text{ex-po}
\end{equation}
The first term $\pick$ represents the direct contribution to the external annihilation. This contribution has the same expression for \Orto and \Para, i.e.~is symmetric with respect to Ps spin configuration, and it is similar to the \quotes{standard} pickoff annihilation rate of Eq.~\myref{pickofffirst}.
To show that, we write $\pick$ separating the spatial and spin part of the Ps wavefunction ($\Psi_{jm}(p,e)=\Psi(\V r _p,\V r _e)\chi_{jm}(\sigma_p,\sigma_e) $). Using the electron density
representation over the single particle spin basis as described in Eqs.~\myref{electronDensity}, \myref{oneBodyRedMproperties1b} and \myref{oneBodyRedMproperties} we get:
\begin{widetext}
\begin{equation}\label{ortoFormal2}
\begin{aligned}
\pick=&\int  |\Psi(\V r _p,\V r _e)|^2n_\uparrow(\V r _1)\left[\chi_{11}(\sigma_p,\sigma_e)s_{\uparrow}(\sigma_1) \hat{\lambda}_1 \chi_{11}(\sigma_p,\sigma_e)s_{\uparrow}(\sigma_1)\right] \\
&+\int  |\Psi(\V r _p,\V r _e)|^2n_\downarrow(\V r _1) \left[\chi_{11}(\sigma_p,\sigma_e)s_{\downarrow}(\sigma_1) \hat{\lambda}_1 \chi_{11}(\sigma_p,\sigma_e)s_{\downarrow}(\sigma_1)\right]  \\
=&\,8\pi a_0^3\int  |\Psi(\V r _p,\V r _e)|^2 \delta(\V r _1 - \V r _p) \left[ \lambda_{3\gamma}n_\uparrow(\V r _1) +\frac{\lambda_{2\gamma}+\lambda_{3\gamma}}{2} n_\downarrow(\V r _1) \right]   \\
=& \, 8\pi a_0^3 \bar{\lambda}\int  |\Psi(\V r _p,\V r _e)|^2n(\V r _p) \\
\end{aligned}
\end{equation}
\end{widetext}
where $\bar{\lambda}$ was defined in Section~\mysec{overcounting} and we have assumed uniform spin distribution of outer electrons, which implies:
\begin{equation}\label{assumptionuniformspin}
n_\uparrow (\V r) =  n_\downarrow(\V r)= \frac{1}{2}n(\V r)
\end{equation}
In the second line, the expectation value of the spin exchange operator inside $\hat{\lambda}_1$ has been obtained expanding the spin part over the eigenstates of $\Sigma_{p,1}$ using the identities:
\begin{equation}\label{identities1}
\begin{aligned}
\chi_{11}(\sigma_p,\sigma_e)s_{\uparrow}(\sigma_1) =& \chi_{11}(\sigma_p,\sigma_1)s_{\uparrow}(\sigma_e)\\
\chi_{11}(\sigma_p,\sigma_e)s_{\downarrow}(\sigma_1) =& \frac{1}{\sqrt{2}}\left[\chi_{00}(\sigma_p,\sigma_1)+\chi_{1,0}(\sigma_p,\sigma_1)\right]s_{\uparrow}(\sigma_e)\\
\chi_{00}(\sigma_p,\sigma_e)s_{\uparrow}(\sigma_1)=& \frac{1}{\sqrt{2}}   \bigg[\chi_{11}(\sigma_p,\sigma_1)s_{\downarrow}(\sigma_e)+ \\
	&\frac{1}{\sqrt{2}}\left(\chi_{00}(\sigma_p,\sigma_1)-\chi_{10}(\sigma_p,\sigma_1)\right)s_{\uparrow}(\sigma_e)\bigg]\\
	\chi_{00}(\sigma_p,\sigma_e)s_{\downarrow}(\sigma_1) =&\frac{1}{\sqrt{2}}\bigg[-\chi_{1-1}(\sigma_p,\sigma_1)s_{\uparrow}(\sigma_e)+\\
	&\frac{1}{\sqrt{2}}\left(\chi_{00}(\sigma_p,\sigma_1)+\chi_{10}(\sigma_p,\sigma_1)\right)s_{\downarrow}(\sigma_e)\bigg]\\
	\end{aligned}
	\end{equation}
The last two identities are written for completeness, because are useful in the analogue calculation on the \Para state.

The second and last integrals $\lambda_\text{ex}$ and  $\lambda_\text{ex-po}$ in Eq.~\myref{generalexpression} are exchange contributions to annihilation.
In $\lambda_\text{ex}$ the annihilation operator directly acts on the Ps spin wavefunction, so that the remaining spin sum is easily performed using the same method of Eq.~\myref{ortoFormal2}. For \Orto it can be shown that the result can be written in a simple form using the definition of the one-body reduced density matrix Eq.~\myref{oneBodyRedM}:
\begin{equation}\label{ortoFormal3a}
\begin{aligned}
\lambda_\text{ex}=&-8\pi a_0^3 \lambda_{3\gamma}\int  \Psi^*(\V r _p,\V r _e)\Psi(\V r _p,\V r _p)\Gamma^{(1)}_{\uparrow\uparrow}(\V r _e;\V r _p)\\
\end{aligned}
\end{equation}
For \Para, $\lambda_\text{ex}$ turns out with the same expression of the above equation after substituting $\lambda_{3\gamma}$ with $\lambda_{2\gamma}$ (this is a consequence of the uniform spin distribution and the fact that $\Gamma^{(1)}_{\uparrow\uparrow}=\Gamma^{(1)}_{\downarrow\downarrow} $).

Finally, the last integral in \myref{generalexpression} is an exchange-correlation contribution to annihilation which can be related to the two-body reduced density matrix $\Gamma^{(2)}$ of the system.
The expectation value of the annihilation operator can be calculated using the spin expansion of $\Gamma^{(2)}$ in Eq.~\myref{twoBodyRedMonSpin}, so that $\lambda_\text{ex-po}$ becomes formally:

\begin{equation}
\begin{aligned}
&\lambda_\text{ex-po} = -2\sum_{ij,i'j'} \sum_{\sigma_e\sigma_p}^{\sigma_1,\sigma_2}\int \diff^3 r_p \diff^3 r_e \diff^3 r_1 \diff^3 r_2  \\
&\qquad\times \Psi^*(\V r _p,\V r _e)\Psi(\V r _p,\V r _2)\Gamma^{(2)}_{iji'j'}( \V r _1,\V r _e;\V r _1,\V r _2)\\
&\quad \times \left[\chi_{11}(\sigma_p,\sigma_e)s_i(\sigma_1)s_j(\sigma_2) \hat{\lambda}_1 \chi_{11}(\sigma_p,\sigma_2)s_{i'}(\sigma_1)s_{j'}(\sigma_e)\right]
\end{aligned}
\end{equation}
After some algebra, using the identities \myref{identities1}, one gets only two non-vanishing contributions for \Orto:
\begin{widetext}
\begin{equation}\label{ortoFormal4a}
\begin{aligned}
 \lambda_\text{ex-po}  = & -2 (8\pi a_0^3) \int \diff^3 r_p \diff^3 r_e
 \diff^3 r_2 \Psi^*(\V r _p,\V r _e)\Psi(\V r _p,\V r _2) \\
& \times\left[  \lambda_{3\gamma}\Gamma^{(2)}_{\uparrow\uparrow\uparrow\uparrow}(\V r _p,\V r _e;\V r _p,\V r _2)
+\frac{\lambda_{2\gamma}+\lambda_{3\gamma}}{2}
\Gamma^{(2)}_{\downarrow\uparrow\downarrow\uparrow}(\V r _p,\V r _e;\V r _p,\V r _2)\right] \\
\end{aligned}
\end{equation}
\end{widetext}

With the same reasoning, but slightly more lengthy calculations, symmetric expressions can be easily obtain for the other \Orto configurations and, in particular, for \Para one obtains:
\begin{widetext}
\begin{equation}\label{ortoFormal4apara}
\begin{aligned}
&\lambda_\text{ex-po} = -2 (8\pi a_0^3) \int \diff^3 r_p \diff^3 r_e \diff^3 r_2 \Psi^*(\V r _p,\V r _e)\Psi(\V r _p,\V r _2) \\
&\times \left[ \frac{\lambda_{2\gamma}+\lambda_{3\gamma}}{2}  \frac{1}{2}\left[ \Gamma^{(2)}_{\uparrow\uparrow\uparrow\uparrow}(\V r _p,\V r _e;\V r _p,\V r _2)+\Gamma^{(2)}_{\downarrow\downarrow\downarrow\downarrow}(\V r _p,\V r _e;\V r _p,\V r _2)\right]\right.\\
& +\frac{\lambda_{2\gamma}-\lambda_{3\gamma}}{2}  \frac{1}{2}\left[\Gamma^{(2)}_{\uparrow\downarrow\downarrow\uparrow}(\V r _p,\V r _e;\V r _p,\V r _2)+\Gamma^{(2)}_{\downarrow\uparrow\uparrow\downarrow}(\V r _p,\V r _e;\V r _p,\V r _2)\right]
 +\left.\lambda_{3\gamma} \frac{1}{2}\left[\Gamma^{(2)}_{\uparrow\downarrow\uparrow\downarrow}(\V r _p,\V r _e;\V r _p,\V r _2)+\Gamma^{(2)}_{\downarrow\uparrow\downarrow\uparrow}(\V r _p,\V r _e;\V r _p,\V r _2)\right]\right]\\
\end{aligned}
\end{equation}
\end{widetext}

Up to this point, the only assumption we made about the system interacting with Ps is that of uniform spin distribution (Eq~\myref{assumptionuniformspin}), a condition which translates in the absence of local spin polarization near the cavity region in the unperturbed ground state of the system.
In particular, no assumption on the form of $\phi$ has been done so that the formulation of the annihilation rate as given in Eq.~\myref{totalannihilation} is completely general.
To provide more physical insight we need to introduce further approximations.

The simplest possible approach is given by the so called local density approximation (LDA).
In LDA, the properties of an electronic system with a density profile $n(\V r)$ are locally modeled at $\V r $ as given by a free electron gas with the same density.
In this simple picture, the 1RDM has an analytical expression~\cite{maruhn2010simple}:
\begin{equation}\label{1RDMFREEheg}
\begin{aligned}
\Gamma^{(1)}(x ; y) &=\delta_{\sigma_x\sigma_y}\frac{n (\V R _{xy})}{2} B\big(k_F(\V R _{xy})\, |\V r _{xy}|\big)\\
\end{aligned}
\end{equation}
where $k_F(\V R _{xy})=\left(3\pi^2 n (\V R _{xy})\right)^{1/3}$ is a \quotes{local} Fermi momentum and
\begin{equation}\label{BR}
\begin{aligned}
B(x)=&3\frac{\sin(x)-x \cos (x)}{x^3}\\
\end{aligned}
\end{equation}
The spatial 1RDM is then:
\begin{equation}
\Gamma^{(1)}(\V R _{xy} ; \V r _{xy})= \Gamma^{(1)}_{\uparrow\uparrow}(\V R _{xy} ; \V r _{xy})+ \Gamma^{(1)}_{\downarrow\downarrow}(\V R _{xy} ; \V r _{xy})
\end{equation}
In Eq.~\myref{1RDMFREEheg} and in the following we use the notation
\begin{equation}
\begin{aligned}
\V R _{xy}& =\frac{\V x + \V y }{2}\\
\V r _{xy}&=\V x - \V y \\
\end{aligned}
\end{equation}
to denote the average and the relative position of two particles $x$ and $y$, respectively.

Whereas the LDA extension of 1RDM is successfully used in standard DFT calculations, a similar result does not hold for 2RDM, which is generally unknown given that it strongly depends on the system under examination.
This is particularly relevant for the calculation of $\lambda_{ex-po}$,  whose terms are proportional to (see Eqs.~\myref{ortoFormal4a} and \myref{ortoFormal4apara}):
\begin{equation}\label{integrationDomain}
\lambda_\text{ex-po} \propto \int \Psi^*(\V r _p,\V r _e)\Psi(\V r _p,\V r _2)\Gamma^{(2)}(\V r _p,\V r _e;\V r _p,\V r _2)
\end{equation}
However we note that, by construction, Ps wavefunctions
$\Psi^*(\V r _p,\V r _e) \Psi(\V r _p,\V r _2)$ exponentially vanish at large interparticle separation, i.e. when $r_{pe},r_{p2} \gtrsim 2 a_0 $ (the Bohr radius for positronium is twice that of hydrogen).
Furthermore, any realistic form of $\Gamma^{(2)}$ should rapidly vanish when inter-particle separation lies in the so called \quotes{exchange-correlation hole} region, whose size is roughly given by the Wigner-Seitz radius $r_s= \left(\frac{3}{4\pi n}\right)^{1/3}$, i.e. the radius of a sphere which on average contains one fermion \cite{wagner2014electron}.
Given that $r_s \gtrsim 2 a_0$ for common values of $n(\V r)$, the integration domain in Eq.~\myref{integrationDomain} is extremely reduced, thus making $\lambda_{ex-po}$ an higher order contribution to the annihilation rate.
For these qualitative reasoning, and given that we are considering only first order corrections to $\lambda$, in the following we will neglect $\lambda_{ex-po}$.

Using the definitions introduced above, the exchange overlap and all the corrections to the annihilation rate can in principle be calculated if the electron density function $n(\V r)$ and the form of Ps spatial wavefunction are known from other computations or other sources.
In the following section we will show how it is possible to include basic qualitative features of these two quantities into the discussion.
However we stress that the theory presented here and in particular Eqs.~\myref{ortoFormal2}, \myref{ortoFormal3a}, \myref{ortoFormal4a} and \myref{ortoFormal4apara} can be evaluated starting from \textit{any} given Ps and electron bulk wavefunctions.

\section{Formal calculation of pickoff annihilation}

In order to find the expression of the spatial Ps wavefunction $\Psi(\V r _p, \V r _e)$, one has to specify the form of its hamiltonian, hence choosing some suitable effective potential $\hat{V}_\text{eff}$ acting on the two particles.
For the sake of simplicity, having in mind a comparison with the TE model, we will focus on a spherical cavity geometry of radius $R_c$  and assume that $\Psi$ can be written in simple factored form using the relative $\V r _{pe} $ and center of mass $\V R _{pe} $ coordinates as:
\begin{equation}\label{Ps-wavefuncion-form}
\Psi(\V r _p, \V r _e)=\psi( r _{pe} )\Psi_\text{TE}( R _{pe})
\end{equation}
Here, the confining effect is taken into account using an infinite potential barrier that keeps the center of mass within a distance $R_c+\Delta$ from the center, where $\Delta$ represents the thickness of the effective interacting region outside $R_c$. Hence the center of mass wavefunction results:
\begin{equation}\label{tewave}
\Psi_\text{TE}(R_{pe}) = \frac{1}{\sqrt{2 \pi (R_c + \Delta)}}\frac{\sin\left(\pi R_{pe}/(R_c+\Delta)\right)}{R_{pe}}
\end{equation}
Since we are neglecting all Coulomb potentials except the one leading to the bound Ps atom, the radial part of the relative wavefunction is supposed to be the same as to the unperturbed Ps, i.e. an Hydrogen-like 1S orbital:
\begin{equation}
\psi( r _{pe} )= \sqrt{k_0}e^{-\frac{r_{pe}  }{2 a_0}}
\end{equation}
We stress again that in place of Eq.~\myref{Ps-wavefuncion-form} one can easily use \textit{any} Ps ground state, obtained from either one-body or two-body models.

On the other side, giving an accurate expression for the electron density function $n(\V r )$ is an extremely complicated task if one has to consider all the interactions naturally present in the system.
Whereas electron-electron repulsion may add a negligible contribution to annihilation, the opposite is true for positron-electron attraction, which would lead to an enhancement of the electron density at the positron position, therefore increasing the annihilation rate.
Without any knowledge of the amount of the enhancement, we can just define a quantity $\rho_e$ to be the effective electron density \textit{felt} by the Ps.
Furthermore, to keep an analogy with TE-like models where the interaction region is limited to a shell layer \footnote[5]{The reader may note that in TE formalism the potential barrier of the confining potential is located at  $R_c +\Delta$ and not at $R_c$, being this last quantity the radius of the free space region.}, we will use:
\begin{equation}\label{n-form}
\begin{aligned}
n(\V r) = \left\lbrace \begin{aligned}
& \rho_e &\qquad\text{if}\qquad r \ge R_c\\
&0 &\qquad\text{if}\qquad  r < R_c\\
\end{aligned}\right.
\end{aligned}
\end{equation}

Using Eqs.~\myref{Ps-wavefuncion-form}, \myref{n-form}, and the LDA expressions \myref{1RDMFREEheg}, the exchange overlap and the symmetric contribution to the annihilation read:
\begin{equation}\label{SAPThegAnnihilation}
\begin{aligned}
&S= \frac{\rho_e}{2}\int_{R_{e1}>R_c} \Psi_\text{TE}( R _{pe})\Psi_\text{TE}( R _{p1})\psi( r _{pe} )\psi( r _{p1} )B(k_F \,r_{e1})\\
&\pick=  \bar{\lambda}\frac{\rho_e}{k_0}\,\int_{r_p >R_c} |\Psi_\text{TE}( R _{pe})|^2 |\psi( r _{pe} )|^2 \\
\end{aligned}
\end{equation}
whereas the exchange correction is given, for \Orto and \Para respectively, by:
\begin{equation}\label{SAPThegAnnihilation_2}
\begin{aligned}
\lambda_\text{ex}^{3\gamma}&= -\frac{\lambda_{3\gamma}\rho_e}{2 \sqrt{k_0}}\int_{R_{pe}>R_c}  \Psi_\text{TE}^*( R _{pe}) \Psi_\text{TE}( r_p)\psi( r _{pe} ) B(k_F \,r_{pe})\\
\lambda_\text{ex}^{2\gamma}&= -\frac{\lambda_{2\gamma}\rho_e}{2 \sqrt{k_0}}\int_{R_{pe}>R_c}  \Psi_\text{TE}^*( R _{pe}) \Psi_\text{TE}( r_p)\psi( r _{pe} ) B(k_F \,r_{pe})\\
\end{aligned}
\end{equation}

Finally, by collecting these first order corrections (as listed in Eq.~\myref{totalannihilation}) and adding the unperturbed intrinsic annihilation, the formal expressions for the total annihilation rates of \Orto and \Para are found:
\begin{equation}\label{final_formal_exact}
\begin{aligned}
\lambda_t &= \left[\lambda_{3\gamma}-\frac{\lambda_\text{ex}^{3\gamma}}{1-S}\right] + \frac{\pick}{1-S}\\
\lambda_s &= \left[\lambda_{2\gamma}-\frac{\lambda_\text{ex}^{2\gamma}}{1-S}\right] + \frac{\pick}{1-S}\\	
\end{aligned}
\end{equation}
These expressions clearly show that a difference in pickoff annihilation rate between Ps states can be ascribed to exchange contributions.

Despite all the approximations used, the integrals appearing in these terms have no analytical expression, so that one still needs to use numerical methods. This is easily done and we will show calculation results in the following section.

However some insights about their qualitative behavior can be deduced using simple geometrical considerations, as follow.
Considering for example the integrand function in the expression of $S$, we note that the radial distances between the three particles $p,e$ and $1$ have a distribution shaped by the exponentials factors $\psi( r _{pe} )\psi( r _{p1} )= \exp[-(r_{pe}+r_{p1})/2a_0]$.
In particular, this means that the integral will be substantially different from zero only when $r _{pe},r_{p1}\lesssim 2 a_0$, i.e. when the two electrons lay altogether around the positron position in a sphere roughly the size of Ps.
Hence, the center of mass positions $R_{pe},R_{p1} $ and $R_{e1} $, which are midway from the corresponding particles, will in turn	lay in a sphere of radius $\approx a_0 $ around $r_p$.
Since this value is generally small compared to the range of variation of $\Psi_\text{TE}( R)$, which in practical cases extends well over the cavity size, we may assume $R_{pe}\sim R_{p1}\sim R_{e1} \sim r_p \equiv R $  and write
\begin{equation}\label{geometrical_approximation}
\Psi_\text{TE}( R _{pe})\Psi_\text{TE}( R _{p1})n(R_{e1}) \approx |\Psi_\text{TE}(R)|^2 n( R)
\end{equation}
Given that $n(R)$ has a step behavior, it's convenient to introduce the quantity $\PoutT$:
\begin{equation}
\PoutT \equiv 4\pi  \int_{R_c}^{R_c+\Delta} |\Psi_\text{TE}(R)|^2 R^2\diff R
\end{equation}
which is the probability of finding the Ps center of mass in the interaction region outside $R_c$, in spherical coordinates.
By using the same approximation to all annihilation contributions, and changing integration variables from $(\V r _p , \V r _i ,\cdots )$ to $(\V R , \V r _{pi} ,\cdots)$, Eqs.~\myref{SAPThegAnnihilation} and \myref{SAPThegAnnihilation_2} become:
\begin{equation}\label{SAPThegAnnihilation2}
\begin{aligned}
S&\approx\frac{\rho_e}{2}\PoutT  \int \psi( r _{pe} )\psi( r _{p1} )B(k_F\,r_{e1})  \diff^3 r _{pe} \diff^3 r _{p1}\\
\pick&\approx  \bar{\lambda}\frac{\rho_e}{k_0} \PoutT \int |\psi( r _{pe} )|^2 \diff^3 r _{pe} = \bar{\lambda}\frac{\rho_e}{k_0}\PoutT \\
\lambda_\text{ex}^{3\gamma}&\approx - \lambda_{3\gamma}\frac{\rho_e}{2 \sqrt{k_0}}\PoutT \int\psi( r _{pe} ) B(k_F\,r_{ep}) \diff^3 r _{pe} \\
\end{aligned}
\end{equation}
and similarly for $\lambda_\text{ex}^{2\gamma}$.
Finally, the total annihilation rates for \Orto and \Para are obtained in the usual form showing the two separate contributions due to the pickoff process:
\begin{equation}\label{final_Orto_app}
\begin{aligned}
\lambda_t &= \left[1- \frac{\PoutT A[\nu]}{1-\PoutT C[\nu]}\right]\lambda_{3\gamma} +  \left[\frac{\rho_e}{k_0}\frac{\PoutT}{1-\PoutT C[\nu]}\right]\bar{\lambda}\\
\lambda_s &= \left[1- \frac{\PoutT A[\nu]}{1-\PoutT C[\nu]}\right]\lambda_{2\gamma} +  \left[\frac{\rho_e}{k_0}\frac{\PoutT}{1-\PoutT C[\nu]}\right]\bar{\lambda}\\
\end{aligned}
\end{equation}
where we have defined the two auxiliary functions $A$ and $C$ by
\begin{equation}
\begin{aligned}
	A[\nu] &=\frac{\rho_e}{2 \sqrt{k_0}}\int\psi( r _{pe} ) B(k_F \,r_{ep}) \diff^3 r _{pe}\\
 C[\nu] &= \frac{\rho_e}{2} \int \psi( r _{pe} )\psi( r _{p1} )B(k_F\,r_{e1})  \diff^3 r _{pe} \diff^3 r _{p1} \\
\end{aligned}
\end{equation}
with $\nu = 2 k_F a_0$, depending on $\rho_e$ through $k_F$.
These functions can be analytically calculated, resulting in:
\begin{equation}
\begin{aligned}
A[\nu] &= \frac{2}{\pi}\left[\arctan(\nu)  -\frac{\nu}{1+\nu^2}\right]\\
C[\nu] &= \frac{2}{\pi}\left[\arctan(\nu)-\frac{\nu-\frac{8}{3}\nu^3-\nu^5}{(1+\nu^2)^3}\right]\\
\end{aligned}
\end{equation}

\begin{figure}
	\centering
	\includegraphics[width=0.9\linewidth]{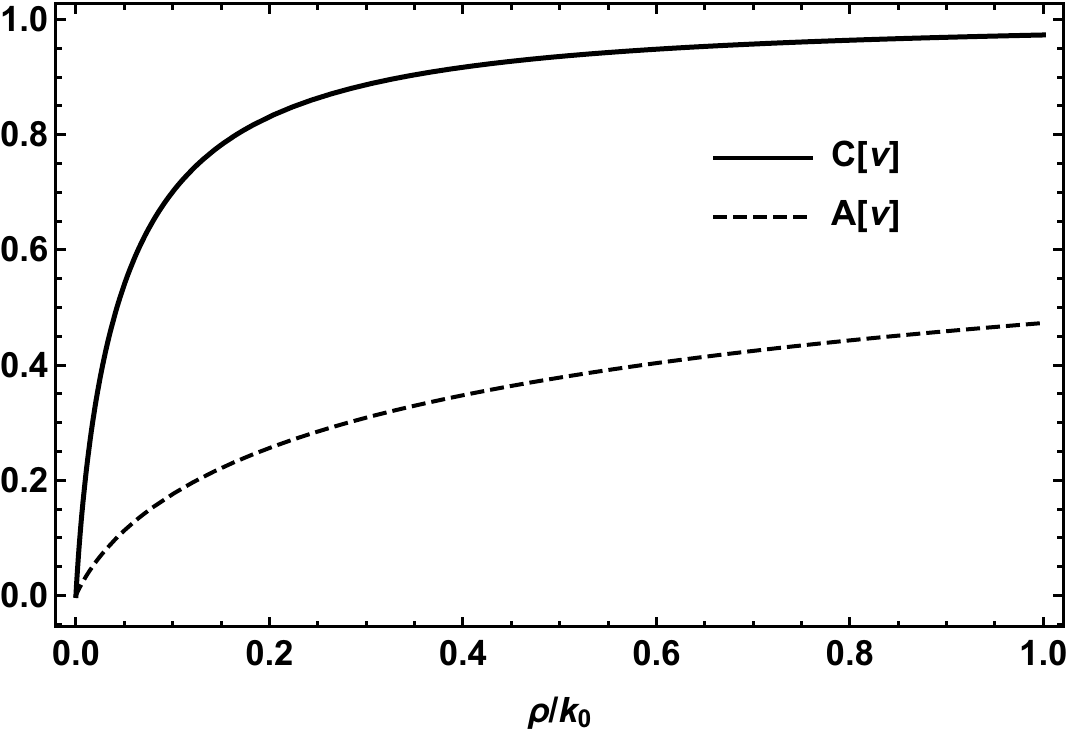}
	\caption{Plot of $C[\nu]$ and $A[\nu]$, with $\nu = 2 k_F a_0$, as a function of the ratio $\rho_e/k_0$, where $ \rho_e$ is the external electron density felt by the positron in the material.  }
	\label{fig:A-C}
\end{figure}
The main advantage of approximation \myref{geometrical_approximation} is that in Eqs.~\myref{final_Orto_app} geometrical effects are well separated from those effects due to electron exchange.
In Fig.~\myfig{fig:A-C} we plot the functions $A$ and $C$, as a function of the (normalized) electron density $\rho_e$ felt by the positron in the material. Both these functions increase for increasing density values while they vanish at the low density limit.

\begin{figure}
	\centering
	\includegraphics[width=\linewidth]{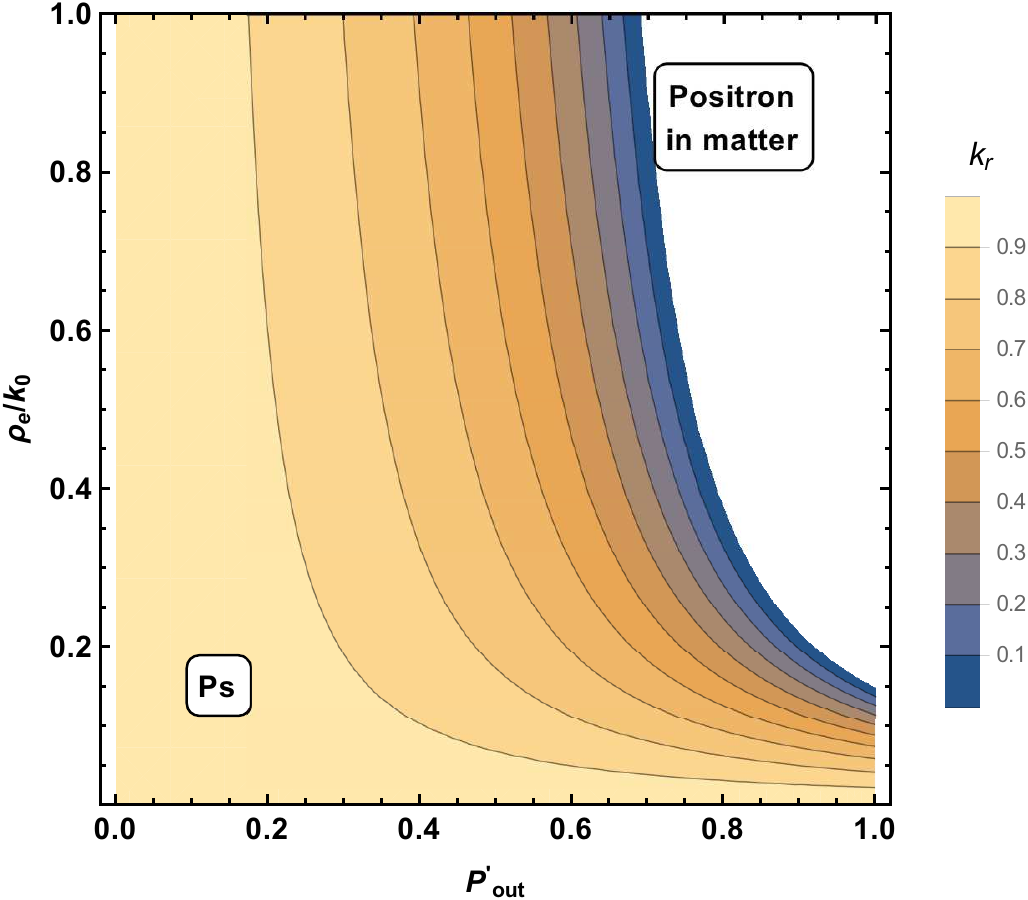}
	\caption{Plot of $k_r$ as a function of the geometrical parameter $\PoutT$ and the electron density $\rho_e$ felt by the positron in the material (Eq.~\myref{kr_approximated}). In the top-right region, $k_r$ assumes negative values since the description of a Ps atom weakly interacting with the environment is no more possible.
	}
	\label{fig:Ps_dissociation}
\end{figure}

Note that within this approximation, the intrinsic relative contact density is given by
\begin{equation}\label{kr_approximated}
k_r = \left[1- \frac{\PoutT A[\nu]}{1-\PoutT C[\nu]}\right]
\end{equation}
By definition, $k_r$ is a useful indicator of the dissociation degree of Ps atom, i.e. of the separability of the Ps-electron.
Its maximum value $k_r= 1$ (Ps in vacuum) is lowered by the overlap with surrounding electrons and vanishes as the original Ps state fades.
When $k_r= 0$ no distinction between \Orto and \Para annihilation rates is possible because all electrons are taken on equal footings. It is important to note that in this picture the vanishing behavior of the contact density is only due to electron indistinguishability and it is by no means related to a spatial deformation of Ps wavefunction, as previously accepted.
In order to show $k_r$ behavior between these two limits, in Fig.~\myfig{fig:Ps_dissociation} we plot its value as a function of both $\PoutT$ and $\rho_e$.

\subsection{Comparison with TE model}
It is interesting to compare Eq.~\myref{final_Orto_app} with the famous Tao-Eldrup result.
Equivalence between the symmetric parts of pickoff annihilation rates predicted by the two models is obtained by setting:
\begin{equation}\label{PoutTEcomparison}
P_\text{out} = \frac{\rho_e}{k_0}\frac{\PoutT}{1-\PoutT C[\nu]}
\end{equation}
Using this scaling condition, we can write Eq.~\myref{final_Orto_app} as:
\begin{equation}\label{final_Orto_app_TE}
\begin{aligned}
\lambda_t &= \left[1- \frac{k_0 }{\rho_e} A[\nu]P_\text{out}\right]\lambda_{3\gamma} +  P_\text{out}\bar{\lambda}\\
\lambda_s &= \left[1- \frac{k_0 }{\rho_e} A[\nu]P_\text{out}\right]\lambda_{2\gamma} +  P_\text{out}\bar{\lambda}\\
\end{aligned}
\end{equation}
which are very similar to Eqs.~\myref{NOovercountedRATES} and of course can be interpreted as in Eqs.~\myref{correct_interpretation}.
In this equivalent version of the TE model, the intrinsic relative contact density in the \textit{surface} region, i.e. when Ps is in the outer shell of thickness $\Delta^\text{TE}$, can be obtained by taking $P_\text{out} = 1$ and turns out to be:
\begin{equation}
k_\text{out}=1-\frac{k_0 }{\rho_e} A[\nu]
\end{equation}
In the limit in which the probability of having an external electron at the positron position reaches the same value of free Ps (i.e. when $\rho_e \to k_0$) $k_\text{out}$ becomes:
\begin{equation}\label{approximation_TE}
\lim_{\rho_e\to k_0}k_\text{out}= 1-0.472 = 0.527
\end{equation}
which is very close to the expectation value of the relative contact density of \textit{one} electron in the negative ion \PSIon \cite{PsIon_Frolov}:
\begin{equation}\label{Ps_ion_contact_density}
\begin{aligned}
\frac{1}{k_0} \bra{ \text{Ps}^- } \hat{\delta} _{p1} \ket{\text{Ps}^- } &= \frac{1}{k_0}\int |\phi_{\text{Ps}^-}(\V r _p,\V r _1,\V r _2)|^2 \delta(\V r _p - \V r _1) \\
&\approx 0.52\\
\end{aligned}
\end{equation}
This analogy is not surprising, given that in the classical picture of $\text{Ps}^-$ only one electron is closely bound to the positron \cite{drachman1982interaction}, and the one-half factor in the contact density comes mainly from the normalization of the total antisymmetric wavefunction.

The result in Eq.~\myref{approximation_TE} can be easily explained with the following simple argument.
Taking as a reference Fig.~\myfig{fig:shielding}, where particles are represented by rigid spheres, we focus on the $m=1$ \Orto, so that both Ps positron and electron will have $\uparrow$ spin configuration.
Then, the pickoff annihilation contribution due only to outer electrons of opposite spin will be proportional to the geometrical probability $ P_{\downarrow}(\V r _p)$ of finding a spin-down electron at the positron position:
\begin{equation}
\Lambda_{\uparrow\downarrow} = P_{\downarrow}(\V r _p) \lambda_{\uparrow\downarrow}
\end{equation}
where $\lambda_{\uparrow\downarrow} $ is the average annihilation rate for opposite-spin configuration.
At the same way, the contribution due to outer electrons of the same spin will be given by the product:
\begin{equation}
\Lambda_{\uparrow\uparrow} = P_{\uparrow}(\V r _p) \lambda_{\uparrow\uparrow}
\end{equation}
From the general expression of the annihilation operator Eq.~\myref{annihilationoperator}, it is easy to find that
\begin{equation}
\begin{aligned}
\lambda_{\uparrow\downarrow} &= \frac{\rho_e}{k_0}\frac{\lambda_{2\gamma} + \lambda_{3\gamma}}{2}; &\lambda_{\uparrow\uparrow} &= \frac{\rho_e}{k_0}\lambda_{3\gamma}\\
\end{aligned}
\end{equation}
where the first expression comes from the fact that the $\uparrow\downarrow$ configuration correspond to a superposition of a $m=0$ \Orto and a \Para. The total annihilation rate then reads:
\begin{equation}
\begin{aligned}
\lambda_t &= \lambda_{3\gamma} + \Lambda_{\uparrow\downarrow} + \Lambda_{\uparrow\uparrow} \\
&= \lambda_{3\gamma} + P_{\downarrow}(\V r _p)\frac{\rho_e}{k_0}\frac{\lambda_{2\gamma} + \lambda_{3\gamma}}{2} + P_{\uparrow}(\V r _p) \frac{\rho_e}{k_0}\lambda_{3\gamma}\\
\end{aligned}
\end{equation}
If no shielding effect is present, and considering uniform spin distribution for outer electrons, $ P_{\downarrow}(\V r _p)$ and $P_{\uparrow}(\V r _p)$ would be equally given by:
\begin{equation}
P_{\downarrow}(\V r _p)=P_{\uparrow}(\V r _p)=\frac{1}{2} P_\text{out}
\end{equation}
where as usual $P_\text{out}$ is the probability of having Ps in the interaction region.
However, the Ps electron tends to \quotes{repel} electrons with the same spin, so that one has $P_{\uparrow}(\V r _p) < \frac{1}{2} P_\text{out} $.
The range of this repulsion is usually associated to the size of the exchange hole, which in turn is inversely proportional to the electron density.
If we assume $\rho_e =k_0$, i.e.~electron density at the positron matching the same value of a $1S$ ground state wavefunction, at most two electrons can be found at the positron position (the Ps electron and an outer one with opposite spin). Hence we have $P_{\uparrow}(\V r _p)=0$ and:
\begin{equation}
\begin{aligned}
\lambda_t &= \lambda_{3\gamma} + \frac{1}{2}P_\text{out}\frac{\lambda_{2\gamma} + \lambda_{3\gamma}}{2}\\
&= (1- \frac{1}{2}P_\text{out} )\lambda_{3\gamma} + P_\text{out}\bar{\lambda}\\
\end{aligned}
\end{equation}
which is just the result of Eq.~\myref{final_Orto_app_TE}.
This suggests a simplified picture in which Ps can be considered as such in the internal cavity region, whereas it resembles a \PSIon \, when inside the interaction region in the external shell.
As a final observation, we note that for lower electron density values, the range of the shielding effect will be wider and in particular for $\rho_e = \rho_0 \approx 0.3 k_0$ it is found that the intrinsic contact density vanishes in the \textit{surface} region:
\begin{equation}\label{maxshielding}
k_\text{out}= \left.1- \frac{k_0 }{\rho_e} A[\nu]\right|_{\rho_e = \rho_0} = 0
\end{equation}
so that Eqs.~\myref{final_Orto_app_TE} become identical to Eq.~\myref{NOovercountedRATES}:
\begin{equation}
\begin{aligned}
\left.\lambda_t\right|_{\rho_0} &= \left[1- P_\text{out}\right]\lambda_{3\gamma} +  P_\text{out}\bar{\lambda}\\
\left.\lambda_s\right|_{\rho_0}  &= \left[1- P_\text{out}\right]\lambda_{2\gamma} +  P_\text{out}\bar{\lambda}\\
\end{aligned}
\end{equation}
then giving a someway stronger justification to the assumptions of the family of models discussed in Section~\mysec{overcounting}.
\begin{figure}
	\centering
\includegraphics[width=\linewidth]{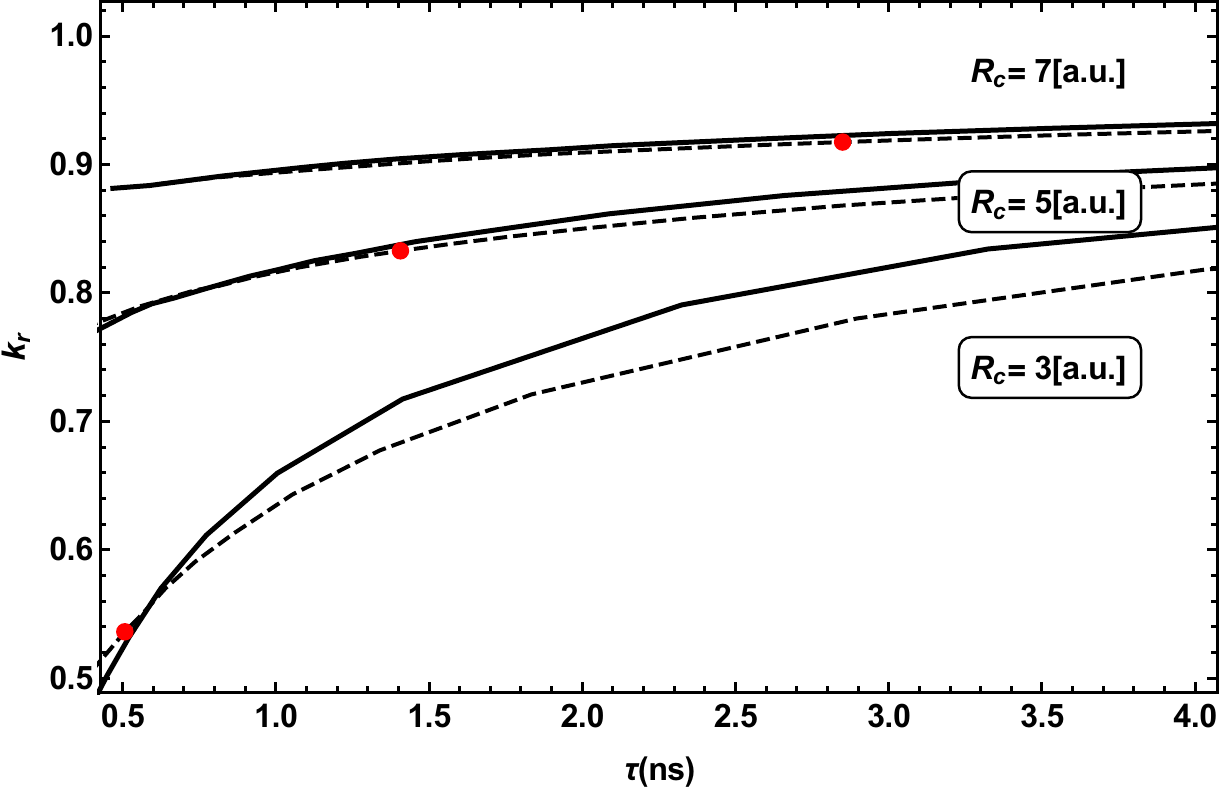}
	\centering
	\caption{Relationship between \Orto lifetime $\tau$ and relative contact density $k_r$, for 3 different values of cavity radius $R_c$.
		The thickness of the interaction layer $\Delta$ was fixed to $3.13$ atomic units to provide comparison with the TE model.
		Curves are obtained by varying the electron density $\rho_e$ felt by Ps. In particular, red points correspond to $\rho_e=k_0$.
		Increasing values of $\rho_e$ correspond to smaller lifetimes and smaller $k_r$. Continuous lines are numerically calculated from Eqs.~\myref{final_formal_exact}, while dotted lines refer to the analytical approximation given in Eqs.~\myref{final_Orto_app}.
		Qualitatively, a lower value of $\Delta$ reproduces the same result of a larger one, if the cavity radius $R_c$ is consequently scaled. }
	\label{fig:tau_kr_D}
\end{figure}

\section{Numerical results}
To better appreciate the role played by geometry in the pickoff annihilation behavior, in Fig.~\myfig{fig:tau_kr_D} we plot the relationship between lifetime $\tau =\lambda_t^{-1}$ and intrinsic relative contact density $k_r$ (Eq.~\myref{kr_approximated}) for a confined \Orto within 3 different choices of the cavity parameter $R_c$. Here, the thickness of the interaction layer was fixed to the TE value $\Delta=3.13\,\text{a.u.}$.
The electron density $\rho_e$ varies in a reasonable range of values and increasing values of $\rho_e$ correspond to shorter lifetime values.
In particular, red points correspond to the choice $\rho_e=k_0$. In these pictures, continuous lines refer to the exact numerical result obtained from Eqs.~\myref{final_formal_exact}, while dashed lines are calculated using the analytical approximation given in Eqs.~\myref{final_Orto_app}.

As expected from the discussion in the previous section, $k_r$ always lie below the vacuum limit $k_r=1$, and gets lower with increasing values of $\Delta$ or $\rho_e$ (i.e. of the overlap $S$) .
It is quite clear that approximations \myref{geometrical_approximation} do not hold for small $R_c$ values, where the heavily distorted wavefunction of the confined Ps undergoes big variations over short distances\footnote[6]{This distortion is to be ascribed to the center of mass motion only, given that there are no potentials acting on the relative part of Ps wavefunction.}.
On the other hand, there is a general good agreement for larger radii.
Also, the variance in predictions between Eqs.~\myref{final_formal_exact} and Eqs.~\myref{final_Orto_app} seems not to be influenced by the value of the electron density, being mainly related to the system geometry.

\begin{figure}
	\centering
\includegraphics[width=1\linewidth]{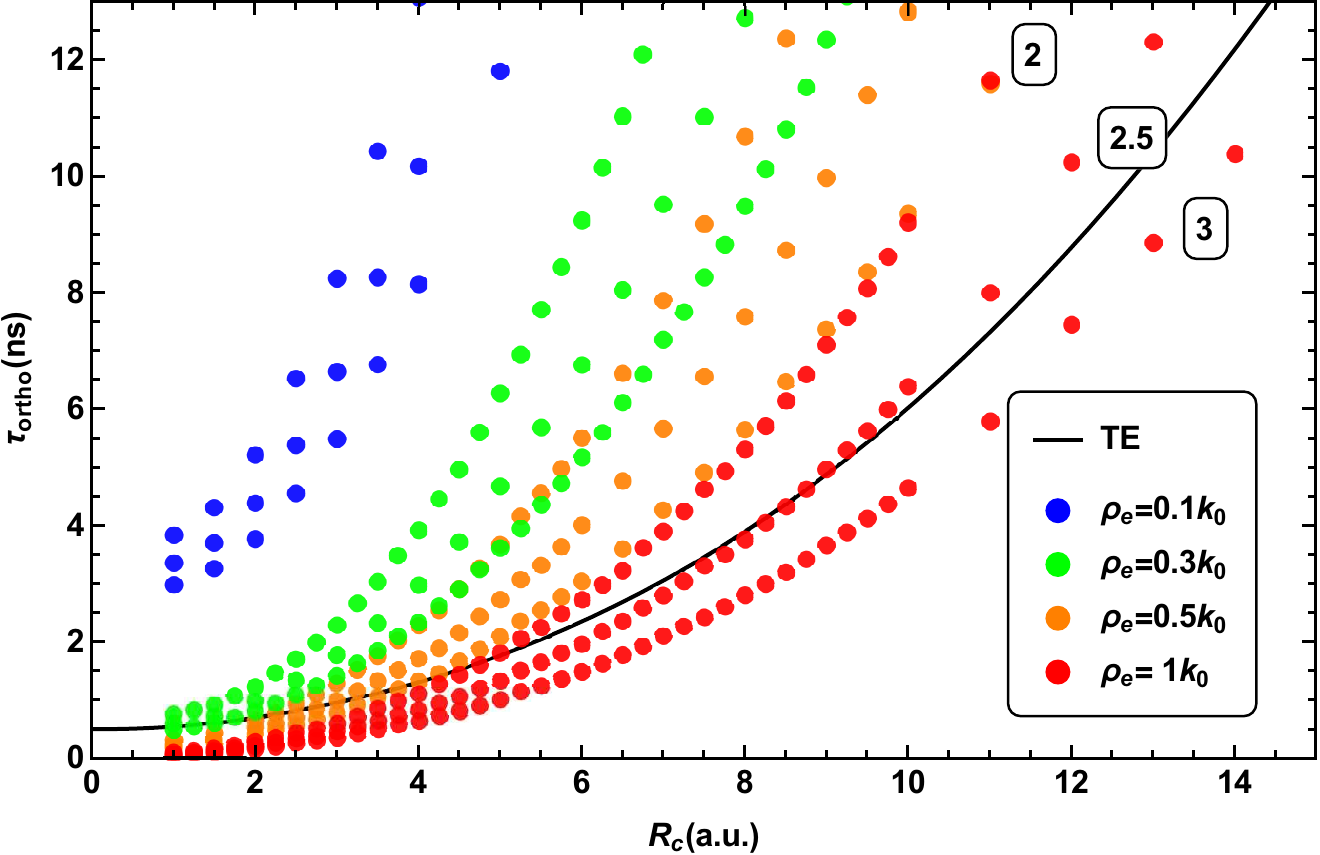}
	\centering
	\caption{Relationship between the cavity radius $R_c$ and \Orto lifetimes, $\tau_\text{Ortho}$, for different values of electron density $\rho_e$ (each represented by a different color).
		Each point was numerically calculated from Eqs.~\myref{final_formal_exact} as a function of $(R_c,\Delta)$, with $\Delta = 2,2.5\text{ and } 3 \,\text{a.u.}$
		Black line corresponds to the TE model prediction, with $\Delta = 3.13 \,\text{a.u.}$
		\label{fig:rc_tau_ortho}
	}
	
\end{figure}

In order to give a proper comparison with TE predictions, Fig.~\myfig{fig:rc_tau_ortho} shows the relationship between the cavity radius $R_c$ and \Orto lifetime, for different values of electron density $\rho_e$ and three choices of $\Delta$.
Here, each color corresponds to a specific value of $\rho_e$.
In particular, red values represent the $\rho_e = k_0$ limit, where the pickoff process can be related to a \textit{surface} formation of \PSIon, as discussed after Eq.~\myref{approximation_TE}.
On the other hand, green values represent the $\rho_e =\rho_0 \approx 0.3 k_0$ limit described in Eqs.~\myref{maxshielding} and \myref{NOovercountedRATES}.
Finally, we also plot an intermediate region $\rho_e = 0.5 k_0$ (in orange) and a low density limit $\rho_e =0.1k_0$ (in blue) for comparison.
The black line represents the TE result \cite{Tao,Eldrup79} and it seems to be compatible with the \PSIon \, formation mechanism (red points), despite the fact that $\Delta$ values considered in this calculation are generally smaller with respect to the commonly accepted TE value.

\begin{figure}
	\centering
\includegraphics[width=1\linewidth]{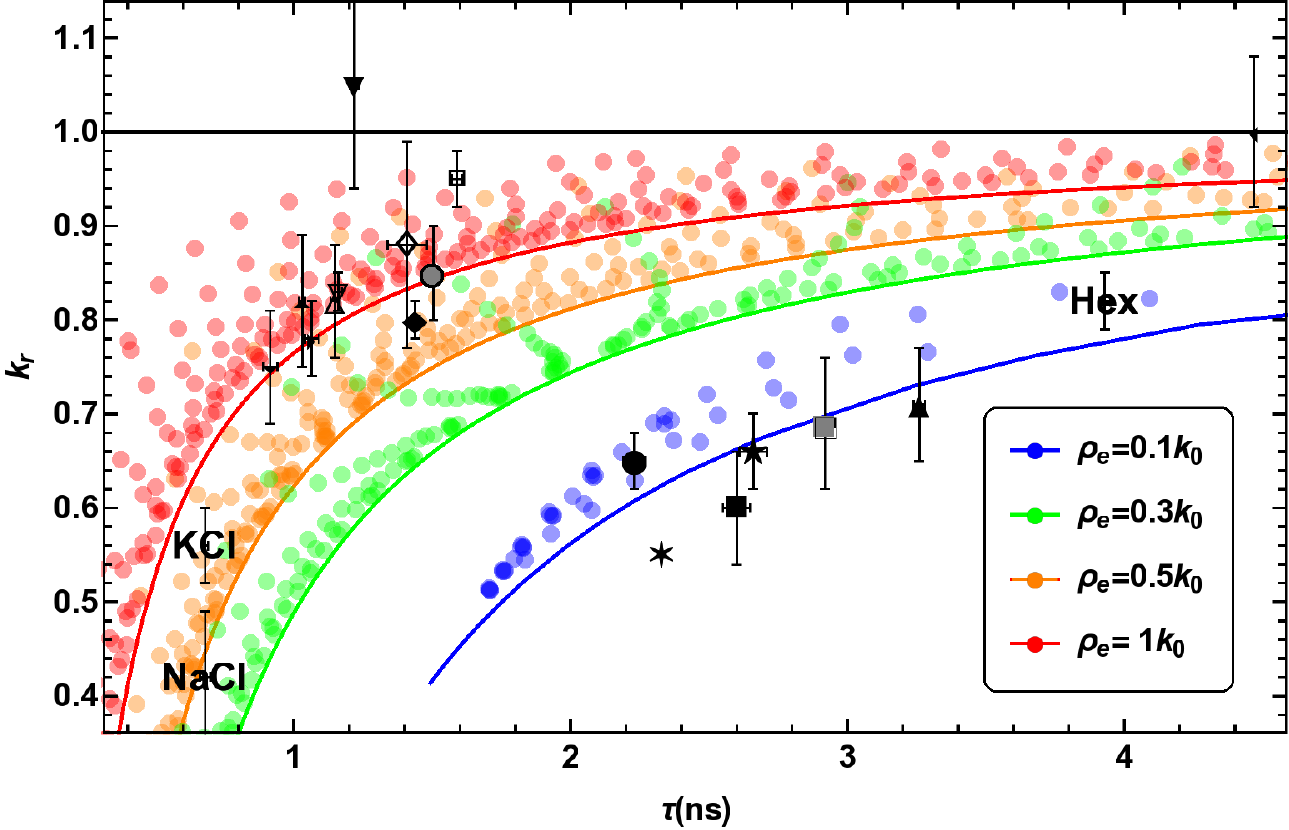}  	
	\centering
	\caption{Relationship between \Orto lifetime $\tau$ and relative contact density $k_r$, for different values of electron density $\rho_e$ (each represented by a different color).
		Each point was numerically calculated from Eqs.~\myref{final_formal_exact} as a function of $(R_c,\Delta)$ (see text), while lines are obtained by the corresponding analytical approximation given by Eqs.~\myref{final_Orto_app}.
		As expected, curves tend to saturate to the analytical approximation in the limit $\Delta \gg R_c$, which corresponds to the situation of a cavity being completely filled with electron gas.
		Known experimental data taken from Table~\ref{Table_data} are plotted for comparison.
		\label{fig:tau_kr_new_01}
	}
	
\end{figure}

In Fig.~\myfig{fig:tau_kr_new_01} we plot the relationship between \Orto lifetime $\tau$ and relative contact density $k_r$, for different values of electron density $\rho_e$, together with known experimental results (see below).
Points are numerically calculated from Eqs.~\myref{final_formal_exact} as a function of the couple $(R_c,\Delta)$, while lines are obtained by the corresponding analytical approximation given by Eqs.~\myref{final_Orto_app}.
The cavity radius $R_c$ and the shell thickness $\Delta$ were taken to vary in the range $1-15\,\text{a.u.}$ and $1-5\,\text{a.u.}$ respectively, a choice in line with the assumption of subnanometric voids.
For comparison, remember that the commonly used TE value of $\Delta$ ($ 3.13\,\text{a.u.}$) was originally obtained from vacancies in the range $R_c\approx 6-8 \,\text{a.u.} $\cite{CanTEbeused}.
As expected, points tend to saturate to the analytical approximation in the limit $\Delta \gg R_c $, which corresponds to the situation of Ps confined in a relatively wide quantum well completely filled with electron gas.

To compare our model with experimental data, we used known results on the contact density $k_r$ and PALS spectra obtained for some polymers and molecular crystals. The data are reported in Table~\ref{Table_data}.
Spectra are decomposed in $3$ or $ 4$ lifetime and relative intensity components.
In the common interpretation, the shorter component $\tau_1\sim 0.125\text{ns}$ is associated to \Para annihilation, the intermediate lifetime $\tau_2 \sim 0.3 \text{ns}$ is due to direct positron annihilation while the longest $\tau_3,\tau_4\sim1-5\text{ns}$ are associated to \Orto annihilating via pickoff process.
For materials having 4 components, points are given in the form $(\tau_4,k_r)$, i.e. as a function of the longest \Orto lifetime component.

Despite this relationship between $\tau_1,\tau_3,\tau_4$ and Ps formation is widely accepted, its implications on the relative intensities $I_1,I_3$ and $I_4$ of the two annihilation channels are rarely taken into account. Indeed, we note that only a few spectra show the correct $I_1/(I_3+I_4)=1/3$ ratio predicted by any model describing \Para/\Orto formation by an unpolarized positron.
This condition could be imposed during the spectrum analysis, but
 it is common practice to ignore it and let all the intensities vary freely during the fitting procedure, thus improving the fit convergence. This because sometimes non-physical values for the lifetimes are obtained by imposing constraints on the intensities.
We want to stress that, without this condition, $\tau_1$ cannot in principle be associated to \Para without introducing arbitrary assumptions on Ps formation mechanism.
This problem may implicate a bias in the estimate of the shorter and longer  components of the spectra (i.e. the one associated to \Para and \Orto respectively).
On the other hand, the relative contact density values $k_r$ in Table~\ref{Table_data} are mostly obtained via magnetic quenching experiments, so that they are largely independent of any possible bias in the PALS analysis.

Fig.~\myfig{fig:tau_kr_new_01} shows a general good agreement between our predictions and a substantial group of the experimental data, which tend to accumulate in the $\rho_e = k_0$ region associated to the \textit{surface} \PSIon \, formation process.
The case of both sodium and potassium chlorides is someway different: they are found in a region characterized by low values of contact density and cavity size.
Here, Ps wavefunction has a high overlap with surrounding electrons, so that its description as a distinct system is blurry.
This is not surprising, given that in such ionic compounds the presence of internal pores or cavity is not expected.
We note also that the other data showing a poor agreement with our model are mostly obtained by a free PALS analysis, that is, without any constraint on the  intensities ratio.
For example, 4 of the 6 points lying in the down-right corner of Fig.~\myfig{fig:tau_kr_new_01} (near the blue curve) present a \Orto / \Para intensity ratio $I_1/I_3 \gtrsim 0.5 > 1/3$. Evidently, when free positrons annihilate in the bulk with a lifetime comparable to that of \Para, it is very hard to disentangle the two components due to the finite resolution of the spectrometer and the interpretation of  $\tau_1$ as pure \Para lifetime is no longer valid.
Thus it is not clear if they are effectively linkable to a Ps trapped in a relatively big cavity ($R_c+\Delta \approx 10,\text{a.u}$ with $ R_c \approx 1\,\text{a.u} $), completely filled with a low density electron gas ($\rho_e \lesssim 0.1k_0$), as would be predicted by the current model.
In particular, the unnaturally high value of $k_r$ found in PPD is associated to an intensity ratio $I_1/I_3 = 4.8 \gg 1/3$ which prevents to identify $\tau_1$ as \Para.

At variance with results obtained with our old model \cite{Tanzi2}, where in the small cavity limit the relative contact density was raised up to the (unphysical) hydrogen value $k_r = 8 $, here $k_r$ tends to vanish.
This different behavior is rapidly explained given the lack, in the current picture, of a confining potential acting on Ps-electron only.
Moreover, the wavefunction describing the electron-positron relative distance inside Ps is exactly the same as in vacuum and the vanishing of the contact density for small values of $R_c$ is a mere consequence of having a higher overlap with outer electrons.

\begin{figure}
	\centering
	\includegraphics[width=1\linewidth]{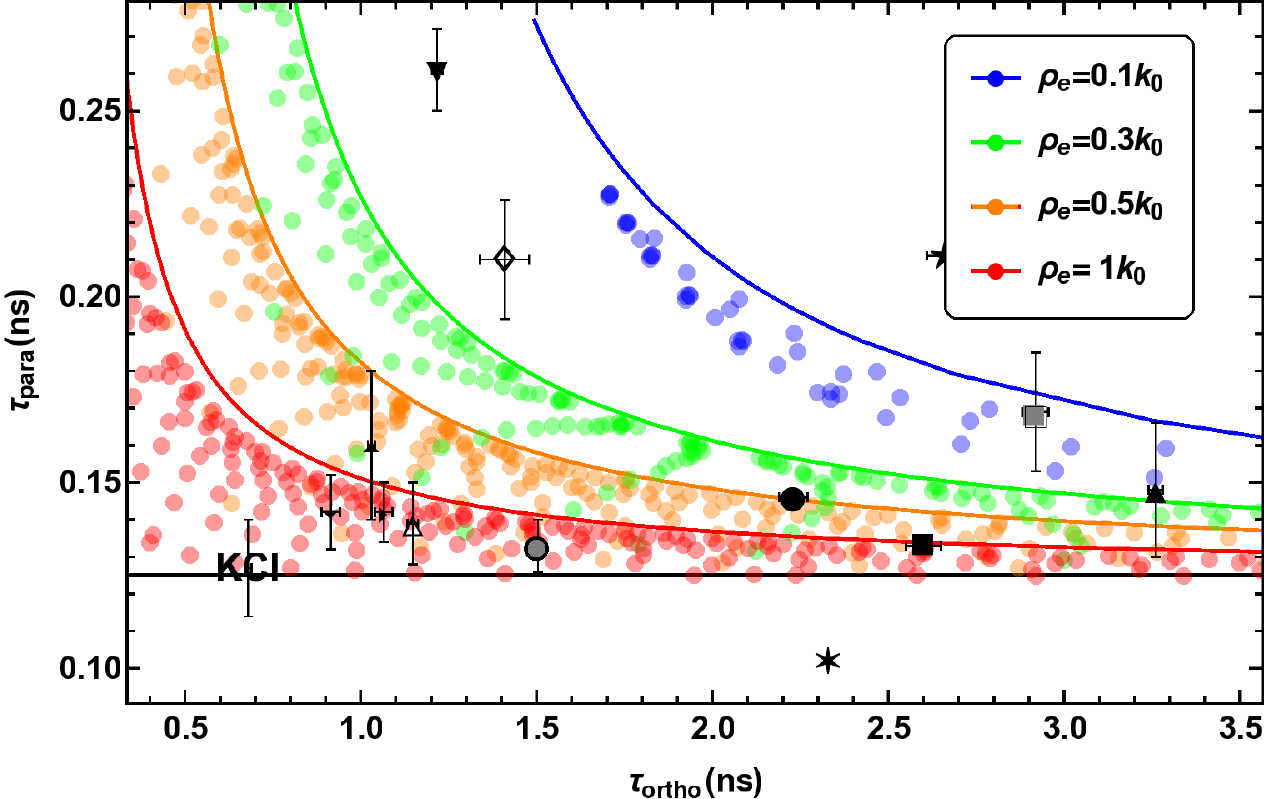}

	\centering
	\caption{Relationship between \Orto and \Para lifetimes, $\tau_\text{Ortho}$ and $\tau_\text{Para}$, for different values of electron density $\rho_e$ (each represented by a different color).
		Each point was numerically calculated from Eqs.~\myref{final_formal_exact} as a function of $(R_c,\Delta)$, while lines are obtained by the corresponding analytical approximation given by Eqs.~\myref{final_Orto_app}.
		As already seen in Fig.~\myfig{fig:tau_kr_new_01}, curves tend to saturate to the analytical approximation at the high $\Delta$ limit, which corresponds to the situation of a cavity being completely filled with electron gas.
		Known experimental data taken from Table.~\ref{Table_data} are plotted for comparison. For materials with more than 3 lifetime components, only the longest one is used.
		The straight black line represents \Para lifetime in vacuum $\lambda_{2\gamma}^{-1} = 0.125\text{ns}$.
		\label{fig:tau_orto_tau_para}
	}
	
\end{figure}

Another useful relationship predicted by our model is that between \Orto and \Para lifetime components, which is plotted in Fig.~\myfig{fig:tau_orto_tau_para} for different values of electron density $\rho_e$.
Like in Fig.~\myfig{fig:tau_kr_new_01}, points are numerically calculated from Eqs.~\myref{final_formal_exact}, while lines are obtained by the corresponding analytical approximation given by Eqs.~\myref{final_Orto_app}.
The straight black line represents \Para lifetime in vacuum $\lambda_{2\gamma}^{-1} = 0.125\text{ns}$.
We can see that most data lay in the range predicted by our model.
Again, the few exceptions (in particular the PPI sample) present a ratio $I_1/I_3$ which does not satisfy the statistical weights $1:3$ of para-to-ortho Ps sublevels.

\begin{sidewaystable}
	\small
	\caption{\label{Table_data} PALS data fo some materials. The relative contact density $k_r$ is mostly obtained by magnetic quenching experiments }
	\begin{tabularx}{\textwidth}{cccccccccccc}
		\hline
Name  & $\tau_1$ & $I_1\%$ & $\tau_2$ & $I_2\%$ & $\tau_3$ & $I_3\%$ & $\tau_4$ & $I_4\%$ & $k_r$ & Sym. & Ref. \\
\hline \hline Atactic polypropylene(PPA)& $0.211 $& $14.3 $& $0.344 $& $54.3 $& \
$0.887 $& $10.2 $& $2.66 \pm 0.05 $& $21.2 $& $0.66 \pm 0.04 $ & \
$\star$ &  \cite{consolati1990origin} \\
Polymethylmethacrylate(PMMA)& $0.146 $& $14.1 $& $0.346 $& $52.9 $& \
$0.875 $& $12.4 $& $2.23 \pm 0.04 $& $20.6 $& $0.65 \pm 0.03 $ & \
$\bullet$ &  \cite{consolati1990origin} \\
Isotactic polypropylene(PPI)& $0.102 $& $14.7 $& $0.334 $& $62.1 $& \
$1.15 $& $9. $& $2.33 $& $14.2 $& $0.55 $ & $\ast$ &  \
\cite{consolati1990origin} \\
Teflon(PTFE)& $0.204 $& $25.2 $& $0.458 $& $46.5 $& $1.58 $& $12.4 $& \
$4.47 $& $15.9 $& $1 \pm 0.08 $ & $\blacktriangleleft$ &  \
\cite{consolati1990origin} \\
Polyethylene(PE)& $0.133 $& $14.7 $& $0.367 $& $59.2 $& $0.989 $& \
$9.9 $& $2.6 \pm 0.05 $& $16.2 $& $0.6 \pm 0.06 $ & $\blacksquare$ &  \
\cite{consolati1990origin} \\
Naphthalene & $0.16 \pm 0.02 $& $18. \pm 3. $& $0.35 \pm 0.01 $& $65. \
\pm 3. $& $1.03 \pm 0.01 $& $17.3 \pm 0.3 $ &  -  &  - & $0.82 \pm \
0.07 $ & $\blacktriangle$ &  \cite{consolati1994magnetic} \\
Acenaphtene& $0.142 \pm 0.01 $& $7.5 \pm 0.4 $& $0.333 \pm 0.004 $& \
$69.9 \pm 1. $& $0.915 \pm 0.026 $& $22.5 \pm 1. $ &  -  &  - & $0.75 \
\pm 0.06 $ & $\blacktriangledown$ &  \cite{consolati1991experimental} \
\\
Byphenil& $0.139 \pm 0.011 $& $7.5 \pm 0.6 $& $0.339 \pm 0.004 $& \
$70. \pm 9. $& $1.148 \pm 0.016 $& $22.5 \pm 0.3 $ &  -  &  - & $0.82 \
\pm 0.06 $ & $\triangle$ &  \cite{consolati1991experimental} \\
KCl& $0.127 \pm 0.013 $& $15.3 \pm 0.5 $& $0.304 \pm 0.004 $& $37.1 \
\pm 2. $& $0.68 \pm 0.01 $& $45.9 \pm 1.5 $ &  -  &  - & $0.56 \pm \
0.04 $ & $\text{KCl}$ &  \cite{consolati1991experimental} \\
Octadecane (solid)& $0.133 \pm 0.007 $& $7.5 \pm 0.2 $& $0.331 \pm \
0.003 $& $70. \pm 0.8 $& $1.504 \pm 0.0015 $& $22.5 \pm 0.6 $ &  -  & \
- & $0.85 \pm 0.05 $ & \textcolor{gray}{$\bullet$} &  \
\cite{consolati1990anomalous} \\
Octadecane (liquid)& $0.169 \pm 0.016 $& $11.4 \pm 0.2 $& $0.463 \pm \
0.031 $& $54.4 \pm 0.7 $& $2.919 \pm 0.036 $& $34.2 \pm 0.6 $ &  -  & \
- & $0.69 \pm 0.07 $ &\textcolor{gray}{ $\blacksquare$ }&  \
\cite{consolati1990anomalous} \\
2,5-Diphenyloxazole (PPO)& $0.142 \pm 0.008 $& $6.8 \pm 0.5 $& \
$0.239 \pm 0.005 $& $32.4 \pm 3. $& $0.44 \pm 0.021 $& $40.3 \pm 1.1 \
$& $1.065 \pm 0.025 $& $20.5 \pm 1.4 $& $0.78 \pm 0.04 $ & \
$\blacktriangleright$ &  \cite{consolati1991positron} \\
2,5-Diphenyl 1,3,4 oxadiazole (PPD)& $0.261 \pm 0.011 $& $53.6 \pm 7. \
$& $0.51 \pm 0.045 $& $35.1 \pm 6. $& $1.217 \pm 0.015 $& $11.2 \pm \
0.4 $ &  -  &  - & $1.05 \pm 0.11 $ & $\blacktriangledown$ &  \
\cite{consolati1991positron} \\
Butyl-PBD& $0.21 \pm 0.016 $& $28.8 \pm 9. $& $0.366 \pm 0.019 $& \
$59.8 \pm 8.5 $& $1.409 \pm 0.07 $& $11.4 \pm 0.4 $ &  -  &  - & \
$0.88 \pm 0.11 $ & $\diamondsuit$ &  \cite{consolati1991positron} \\
p--terphenyl (doped with anthracene) &  -  &  - & $0.313 \pm 0.004 $ & \
- & $1.438 \pm 0.015 $& $19.9 \pm 0.2 $ &  -  &  - & $0.8 \pm 0.02 $ \
& $\blacklozenge$ &  \cite{goworek1994normal} \\
p--terphenyl (doped with chrysene) &  -  &  - & $0.317 \pm 0.003 $ &  \
- & $1.16 \pm 0.013 $& $22.5 \pm 0.3 $ &  -  &  - & $0.83 \pm 0.02 $ \
& $\triangledown$ &  \cite{goworek1994normal} \\
NaCl &  -  &  -  &  -  &  - & $0.68 \pm 0.02 $ &  -  &  -  &  - & \
$0.42 \pm 0.07 $ & $\text{NaCl}$ &  \cite{bisi1973properties} \\
Hexane (degassed)& $0.29 \pm 0.02 $& $40. \pm 3. $& $0.86 \pm 0.04 $& \
$16. \pm 2. $& $3.93 \pm 0.02 $& $33. \pm 1. $ &  -  &  - & $0.82 \pm \
0.03 $ & $\text{Hex}$ &  \cite{bisi1982positronium} \\
Benzene& $0.148 \pm 0.018 $& $18.1 \pm 2.2 $& $0.416 \pm 0.007 $& \
$37.4 \pm 1. $& $1.18 \pm 0.18 $& $5.3 \pm 0.8 $& $3.26 \pm 0.02 $& \
$39.2 \pm 0.8 $& $0.71 \pm 0.06 $ & $\blacktriangle$ &  \
\cite{consolati1991unusual} \\
Alpha-SiO2 &  -  &  -  &  -  &  - & $0.27 \pm 0.01 $ &  -  &  -  &  - \
& $0.31 \pm 0.02 $ & $\spadesuit$ &  \cite{nagai1999lifetime} \\
Amorphous-SiO2 &  -  &  -  &  -  &  - & $1.59 \pm 0.02 $ &  -  &  -  &  - & \
$0.95 \pm 0.03 $ & $\square$ &  \cite{nagashima2001positronium} \\
		\hline
	\end{tabularx}
	
\end{sidewaystable}	

\section{Conclusions}

The description of annihilation behavior of Ps atoms in
nanoporous materials has been addressed in literature by
means of various theoretical frameworks, based both on
one-body or two-body basic models. Among these attempts,
only in a few cases attention was given to the unavoidable
presence of exchange effects between Ps-electron and outer
electrons. These effects, supposedly, can affect in some
relevant extent the pickoff annihilation, and pose the question
if Ps can be effectively seen as a separate "entity"
where the Ps electron is somehow privileged with respect to
outer electrons.

In this paper we face this problem using
symmetry adapted perturbation theory (SAPT), managing to
set up a theoretical framework to formally calculate Ps
annihilation rates in realistic material conditions.
With the help of the analysis developed here, we were able to
clarify some concepts that had had many different interpretations in
literature.
In particular, we managed to provide insights about the meaning of the relative contact density $k_r$, which for long time has been related only to the spatial part of the confined Ps wavefunction.
Also, we clarify the form of the pickoff term describing the annihilation process of Ps in cavities, which has been always
taken to be identical for \Orto and \Para. Furthermore, we focused
on a particular aspect of this problem, present on simple descriptions
of pickoff processes, which we call \quotes{over-counting}.

Using a simplified model of a Ps interacting with an N-electrons environment, we showed how the pickoff annihilation rate is indeed different for \Orto and \Para.
In practice, we found that a spin-shielding effect must be ascribed to the Ps-electron, which makes the pickoff process asymmetric with respect to the two Ps spin configurations, a feature often misunderstood and never previously analyzed in literature. On the other hand, it is possible to reconnect known results with ours by recasting this difference in a symmetric form directly related to the observed lowering of the intrinsic contact density, hence with a parallel and new interpretation of the whole annihilation processes.
Indeed, within SAPT framework, $k_r$ essentially becomes an indicator of the dissociation degree of Ps atom, i.e. of the separability of the Ps-electron with respect to other electrons of the surrounding.
Its maximum value $k_r= 1$ (Ps in vacuum) is lowered by the overlap with surrounding electrons and it vanishes as the original Ps state fades.
When $k_r= 0$ no distinction between \Orto and \Para annihilation rates is possible because all electrons are taken on equal footings.

In particular, we suggested a new model which only depends on 3 parameters, namely the size $R_c$ of the free space region (cavity), the thickness of the interaction layer $\Delta$ and the value of the outer electron density $\rho_e$ effectively interacting with the positron.
Finally, our main result can be summarized with the expressions of the total annihilation rate of \Orto and \Para (Eqs.~\myref{final_formal_exact}).

Remarkably, we found that our model is capable to provide a simple explanation for the lowering of the contact density, despite it is characterized by the complete lack of \textit{any} potential that could pull the electron and positron apart.
Indeed, we used an expression for the Ps relative wavefunction which is exactly the same as in vacuum.
The vanishing of the contact density for small values of $R_c$ or for high values of $\rho_e$ is then a mere consequence of having an higher overlap with outer electrons, and it is by no means related to a spatial deformation of Ps wavefunction, as previously believed and accepted.
In other words, it is the concept of Ps itself which inevitably fades out when electrons can no more be distinguished.
As a final remark, we note that, despite we have used a simple form of the Ps wavefunction, our results can be easily extended to any one-body and two-body model describing Ps in matter.

Further investigations are necessary to test and validate the relationships provided by our model.
In particular, PALS and magnetic quenching experiments on materials subjected to external pressure, as for example~\cite{zanatta2014structural}, can be extremely useful as they reduce the number of unknown free parameters.
As also recommended above, to avoid interpretation bias future PALS analysis should be performed assuming the correct intensity ratio between \Para and \Orto lifetime components, a condition which can be easily achieved by a constrained fitting procedure.

\section*{Acknowledgements}
Work done in partial fulfillment of the requirements for the Ph.D.
degree in Physics by G. Tanzi Marlotti at the Universit\`{a} degli Studi di Milano.

%

\end{document}